# Synthesis and characterization of sodium-iron antimonate Na$_2$FeSbO$_5$: One-dimensional antiferromagnetic chain compound with spin-glass ground state

*Sitharaman Uma[1,#], Tatyana Vasilchikova[2], Alexey Sobolev[3], Grigory Raganyan[2], Aanchal Sethi[1], Hyun-Joo Koo[4], Myung-Hwan Whangbo[5,6,7], Igor Presniakov[3], Iana Glazkova[3], Alexander Vasiliev[2,8,9], Sergey Streltsov[10,11] and Elena Zvereva[2*]*

[1]Materials Chemistry Group, Department of Chemistry, University of Delhi, Delhi-110007, India

[#]sitharamanuma@yahoo.com

[2]Physics Faculty, M.V. Lomonosov Moscow State University, Moscow 119991, Russia

*zvereva@mig.phys.msu.ru

[3]Chemistry Faculty, M.V. Lomonosov Moscow State University, Moscow 119991, Russia

[4]Department of Chemistry and Research Institute for Basic Sciences, Kyung Hee University, Seoul 130-701, Korea

[5]Department of Chemistry, North Carolina State University, Raleigh, NC 27695-8204, USA

[6]State Key Laboratory of Crystal Materials, Shandong University, Jinan 250100, China

[7]State Key Laboratory of Structural Chemistry, Fujian Institute of Research on the Structure of Matter (FJIRSM), Chinese Academy of Sciences (CAS), Fuzhou 350002, China

[8]National Research South Ural State University, Chelyabinsk 454080, Russia

[9]National University of Science and Technology "MISiS", Moscow 119049, Russia

[10]Institute of Metal Physics RAS, Ekaterinburg 620990, Russia

[11]Ural Federal University, Ekaterinburg 620002, Russia

**ABSTRACT**

A new oxide, sodium–iron antimonate $Na_2FeSbO_5$, was synthesized and structurally characterized, and its static and dynamic magnetic properties were comprehensively studied both experimentally by *dc* and *ac* magnetic susceptibility, magnetization, specific heat, ESR and Mössbauer measurements and theoretically by density functional calculations. The resulting single crystal structure ($a$ = 15.6991(9) Å; $b$ = 5.3323 (4) Å; $c$ = 10.8875(6) Å; S.G. *Pbna*) consists of edge-shared $SbO_6$ octahedral chains, which alternate with vertex-linked, magnetically active $FeO_4$ tetrahedral chains. The $^{57}$Fe Mössbauer spectra confirmed the presence of high spin $Fe^{3+}$ ($3d^5$) ions in distorted tetrahedral oxygen coordination. The magnetic susceptibility and specific heat data show absence of a long-range magnetic ordering in $Na_2FeSbO_5$ down to 2 K, but *ac* magnetic susceptibility unambigously demonstrates spin-glass-type behavior with unique two-step freezing at $T_{f1} \approx 80$ K and $T_{f2} \approx 35$ K. Magnetic hyperfine splitting of $^{57}$Fe Mössbauer spectra was observed below $T^* \approx 104$ K ($T_{f1} < T^*$). The spectra just below $T^*$ ($T_{f1} < T < T^*$) exhibit a relaxation behavior caused by critical spin fluctuations indicating the existence of short-range correlations. The stochastic model of ionic spin relaxation was used to account for the shape of the Mössbauer spectra below the freezing temperature. A complex slow dynamics is further supported by ESR data revealing two different absorption modes presumably related to ordered and disordered segments of spin chains. The data imply a spin-cluster ground state for $Na_2FeSbO_5$.

**KEYWORDS**: Spin 5/2 chain one-dimensional; Single crystal XRD; ESR; Mössbauer spectroscopy; Density functional calculations

## 1. INTRODUCTION

The design and optimization of materials for the purpose of identification of new electrodes for Li (Na) batteries is an ongoing challenging area of research. Among many systems that are being investigated, there are quite a few compounds hosting Li$^+$ or Na$^+$ ions in combination with transition metal cations, particularly with Fe or Mn. The notable examples are LiCoO$_2$, LiNi$_{1/3}$Mn$_{1/3}$Co$_{1/3}$O$_2$ belonging to the well-established AMO$_2$ (A = Li, Na; M = transition, post transition element or their combination) family of oxides, together with LiMn$_2$O$_4$ and the olivine-type compounds like LiFePO$_4$.[1] The selection of materials for designing the electrodes is a challenging task that requires the consideration of various factors including the structure, the type and potential of redox metal ions, the composition and morphology, etc. Choosing an appropriate synthetic method plays a key role in the entire process. One specific direction for synthetic exploration, which is actively developing during two last decades, has been the introduction of multiple metal ions (2/3 of M$^{2+}$ ions with 1/3 of M$^{5+}$ ions) replacing M$^{3+}$ ions in AMO$_2$ stoichiometry. The resulting oxides belong to the ordered rocksalt superstructure variants. The various members of this family, in particular Li$_3$M$_2$XO$_6$ (M = Mg, Co, Ni, Cu, Zn; X = Nb, Ta, Sb, Bi)[2-6] and Na$_3$M$_2$XO$_6$ (M = Mg, Co, Ni, Cu, Zn; X = Sb, Bi)[7-10] excluding those containing niobium and tantalum have edge shared honeycomb ordered layers (M$_2$XO$_6$)$^{3-}$ separated by Li$^+$/Na$^+$ interlayer ions. Additional examples include Li$_4$MTeO$_6$ (M = Co, Ni, Cu, Zn)[11-13] and Li$_4$MSbO$_6$ (M = Cr, Fe, Al, Ga, Mn)[11,14-15] along with Na$_4$FeSbO$_6$[16,17] and Na$_3$LiFeSbO$_6$.[18] The latter oxides contain twice less of magnetic ions and as a consequence triangular magnetic sublattice. Out of these oxides, Na$_3$Ni$_2$SbO$_6$, and Li$_4$NiTeO$_6$ have been investigated as potential cathode materials for Na$^+$-ion and Li$^+$-ion batteries.[19,20] Li$_4$FeSbO$_6$ has been shown to exhibit a complex electrochemical behaviour; the oxidation of iron and oxygen together occurs at 4.2 V during charging with the release of oxygen at higher voltages.[15,21]

The honeycomb topology of the magnetic $M^{2+}$ ion lattices found in $Li_3M_2XO_6$ and $Na_3M_2XO_6$ and other related oxides is responsible for the plenty of unusual magnetic phenomena and stabilization of exotic quantum states in this class of materials. For example, a spin-gap behaviour has been reported in $Na_2Cu_2TeO_6$,[22] $Na_3Cu_2SbO_6$[23,24] and the presence of fragmented Haldane-like chains in $Li_3Cu_2SbO_6$.[25] An antiferromagnetic ordering at low temperatures has been observed for $Na_3M_2SbO_6$ (M = Ni, Co),[8,9,26,27] $Na_{3.70}Co_{1.15}TeO_6$,[28] $Li_3Ni_2SbO_6$,[26,29,30] $Ag_3Co_2SbO_6$,[31] and in $A_3Ni_2BiO_6$ (A = Li, Na).[6,10] Other oxides, $Na_4FeSbO_6$, $Na_3LiFeSbO_6$ and $Li_4MSbO_6$, demonstrated the absence of a long range magnetic order.[14-18] Several other rock-salt based oxides, $Li_4M^{II}TeO_6$ (M = Co, Ni, Cu, Zn),[11-13,15] $Li_4M^{III}SbO_6$ (M = Al, Cr, Fe),[11,15,16] $Li_3(Li_{1.5x}Fe_{3-(x+1.5x)})TeO_6$ ($0.1 \leq x \leq 1.0$)[32] and $Li_4M^{III}_{0.5}Te_{1.5}O_6$ (M = Cr, Mn, Al, Ga)[33] have been structurally characterized, but investigations of their magnetic properties are limited at the moment.

Subsolidus phase relations in $Na_2O-Fe_2O_3-Sb_2O_x$ have been studied earlier by Politaev and Nalbandyan,[16] who briefly reported the formation of $Na_2FeSbO_5$ with (S.G *Pbcn*; *a* = 10.8965(13) Å; *b* = 15.7178(13) Å; *c* = 5.3253(4) Å) based on the indexation of the powder X-ray diffraction (PXRD) pattern. They noted similarity with brownmillerite structure in lattice metrics and coordination of the cations but considerable difference in diffraction intensities. In the present work, we report a successful structure determination of $Na_2FeSbO_5$ using single crystals obtained from sodium sulfate flux. The structure solution substantiated the observed PXRD pattern for the bulk polycrystalline sample of $Na_2FeSbO_5$. The resulting structure confirms the space group (No. 60) and predicted coordination numbers based on the formula volume;[16] six for Sb and Na and four for Fe. It differs from the brownmillerite structure but still possesses the corner-linked chains of $FeO_4$ tetrahedra along the shortest unit-cell edge. Mixed metal oxides of $Fe^{3+}$ and $Sb^{5+}$ are known to crystallize in different structures such as rutile ($FeSbO_4$),[34] perovskite ($Ba_2FeSbO_6$)[35] and pyrochlore ($Pr_2FeSbO_7$, $Nd_{1.8}Fe_{0.2}(FeSb)O_7$)[36] and to exhibit a spin-glass ground state. In contrast in the crystal structure of $Na_2FeSbO_5$, there are edge shared zigzag octahedral ($SbO_6$) chains linked through corners to tetrahedral ($FeO_4$) chains formed by vertex oxygen sharing. This type of structure provides the conditions for low-dimensional spin-exchange interactions and frustration in magnetic sublattice. In what follows, we report the synthesis of $Na_2FeSbO_5$, the characterization of its structure and detailed study of its magnetic properties.

## 2. EXPERIMENTAL SECTION

### 2.1. Synthesis

Polycrystalline $Na_2FeSbO_5$ was prepared by standard solid-state method. A mixture of $Na_2CO_3$ (Fischer Scientific, 99.9 %), $Fe_2O_3$ (Sigma Aldrich, ≥ 99.0 %) and $Sb_2O_3$ (Sigma Aldrich, + 99.9 %) in the ratio of 1:0.5:0.5 was thoroughly ground and heated in an alumina crucible to 700 °C followed by 1000 °C for 12 h with several intermittent grindings. Crystals of $Na_2FeSbO_5$ were grown using $Na_2SO_4$ flux. Stoichiometric amounts of $Na_2CO_3$, $Fe_2O_3$ and $Sb_2O_3$ were ground with excess of $Na_2SO_4$ (5 times by weight) and heated at 1000 °C for 3 h in an alumina crucible. This was followed by slow cooling (1 °C/h) to 950 °C, and then to 900 °C with a rate of 2 °C/h and further down to 800 °C with a rate of 5 °C/h and finally the furnace was cooled to room temperature. Single crystals with light brown colour obtained were washed with hexane and were used for further single crystal X-ray diffraction (XRD) measurements.

### 2.2. Characterization

Energy Dispersive X-ray (EDX) measurements of solid samples were carried out using JEOL 6610 LV scanning electron microscope. Field emission scanning electron microscopy (FE-SEM) images of the samples were collected using a Carl Zeiss Gemini SEM 500 microscope. Single crystal X-ray diffraction data of freshly grown single crystals were collected using an Oxford Xcalibur NOVA diffractometer with a four circle κ goniometer employing a graphite-monochromatized Mo Kα (λ = 0.71073 Å) radiation at 150 K. The diffraction intensities were corrected for Lorentz and polarization effects. The data were reduced using CrysAlisRED[37] (programs available with the diffractometer), the shape and size of the crystal were determined with the video microscope attached to the diffractometer, and an analytical absorption correction (after Clark and Reid) from the crystal shape was applied.[38] The crystal structure was determined and refined by direct methods using SHELXS-97 incorporated in WINGX suite.[39,40] All atoms in the structure were refined using full matrix least-squares methods on $F^2$.

The EDX analysis of the single crystals obtained from the $Na_2SO_4$ flux and that of the polycrystalline sample confirmed the Na:Fe:Sb atomic ratio of 2:1:1 (Figure S1, Supporting Information). The preliminary experiments in the single crystal X-ray diffractometer resulted in orthorhombic lattice parameters, $a$ = 15.6991(9) Å, $b$ = 5.3323(4) Å, $c$ = 10.8875(6) Å. The systematic absences suggested the *Pbna* space group. The solution using direct methods resulted in locating two antimony, one iron and three sodium atom positions in the asymmetric unit. Subsequent refinement cycles assisted in locating the five oxygen atoms, thus yielding the expected stoichiometry, $Na_2FeSbO_5$. Refinements of the site occupancies did not show significant deviations. The anisotropic displacement parameters were obtained for all the atoms other than the oxygen. The largest residual electronic density peak and hole in the final difference map were +7.85 and -2.915 e/Å$^3$. The crystallographic data, positional and thermal parameters are summarized in Tables 1-2, with the anisotropic displacement parameters listed in Table S1. The packing diagrams were generated by DIAMOND version 3.[41]

PXRD measurements were taken at room temperature on a PANalyticalX'Pert PRO diffractometer equipped with Cu K$α$ radiation ($\lambda$ = 1.5418 Å). Data were collected in the angular range of 2θ = 3 – 70° with a scan step width of 0.04° and a scan rate of 4.5 s/step. Using these data, Le Bail fit was carried using the TOPAS 3 software, which estimates the background using a Chebyshev polynomial function with 5 coefficients, and describes the peak shape by a pseudo-Voigt function.[42] The zero error, shape parameters, lattice parameters and the profile coefficients were refined to obtain a suitable fit.

Second Harmonic Generation (SHG) measurements on powder samples were performed by the Kurtz powder technique. In this experiment, Q-switched pulses were obtained from a Nd-YAG laser of wavelength 1064 nm with pulse duration of 10 ns and frequency repetition of 10 Hz, which was passed through the sample powder to get maximum SHG efficiency. The sample was ground to a particle size of 63 μm, packed in a micro capillary of uniform bore and then exposed to laser radiation. The second harmonic radiations generated by the randomly oriented micro crystals were focused by a lens and detected by a photomultiplier tube and then converted into electrical signal. The signal amplitude in volts indicated the SHG efficiency of the sample. KDP crystals grounded into identical size (63 μm) were used as the reference material.

Mössbauer (MS) experiments were performed in the temperature range between 17 - 300 K in closed-cycled Janis CCS-850-1 cryostat in transmission geometry with a 900 MBq $\gamma$-source of $^{57}$Co(Rh) mounted on a conventional constant acceleration drive. The radiation source $^{57}$Co(Rh) was kept at room temperature. All isomer shifts refer to the α-Fe absorber at 300 K. The experimental Mössbauer spectra were analyzed using the *SpectrRelax* software package.[43]

The magnetic measurements were performed by means of a Quantum Design PPMS system using teflon capsule. The temperature dependence of the *dc* magnetic susceptibility was measured at the magnetic field $B$ = 0.1, 1 and 9 T in the temperature range 1.8 – 400 K. The temperature dependence of the *ac* magnetic susceptibility was measured at the magnetic field $B$ = 0.0001 T in the temperature range between 2 – 300 K while varying the frequency $f$ between $f$ = 0.5 – 10 kHz. The isothermal magnetization curves were obtained for magnetic fields $B \leq 7$ T at $T = 1.8, 2.5, 5, 50, 70, 90, 200$ K after cooling the sample in zero magnetic field.

Specific heat measurements were carried out by a relaxation method using a Quantum Design PPMS system on the cold-pressed $Na_2FeSbO_5$ sample. The data were collected at zero magnetic field and under applied field of 9 T in the temperature range $2 - 300$ K.

Electron spin resonance (ESR) studies were carried out using an X-band ESR spectrometer CMS 8400 (ADANI) ($f \approx 9.4$ GHz, $B \leq 0.7$ T) equipped with a low-temperature mount, operating in the range $T = 6 - 420$ K. The effective g-factor of our sample has been calculated with respect to an external reference for the resonance field. We used BDPA (a,g - bisdiphenyline-b-phenylallyl) $g_{et} = 2.00359$, as a reference material.

## 3. RESULTS AND DISCUSSION

### 3.1. Synthesis

The bulk, polycrystalline $Na_2FeSbO_5$ can readily be synthesized in air at 1000 °C and was found to remain stable without any noticeable decomposition. The PXRD pattern matched well with the reported pattern indicating an orthorhombic lattice ($a \approx 10.896$ Å; $b \approx 15.718$ Å; $c \approx 5.325$ Å) resembling a brownmillerite structure (Figure 1). Commonly, this structure with composition, $Ca_2FeAlO_5$, which many oxides including $Ca_2Fe_2O_5$ adopt has been known to crystallize in orthorhombic symmetry with unit cell dimensions, $\sqrt{2}a_p \times 4a_p \times \sqrt{2}a_p$ ($a_p$ = perovskite unit cell parameter of $\sim 4$ Å).[44]

This structure is built up of alternating layers of vertex-linked $MO_6$ (M = metal) octahedra and $MO_4$ tetrahedra. Several structural variations are possible based on how the chains of corner-sharing $MO_4$ tetrahedra are ordered. Accordingly, brownmillerites can have a range of structures with different lattice parameters and space groups.[45] A brownmillerite superstructure $Ca_2FeCoO_5$ (S.G. Pbcm) has the unit cell dimensions of $a = 5.3652(3)$ Å; $b =$

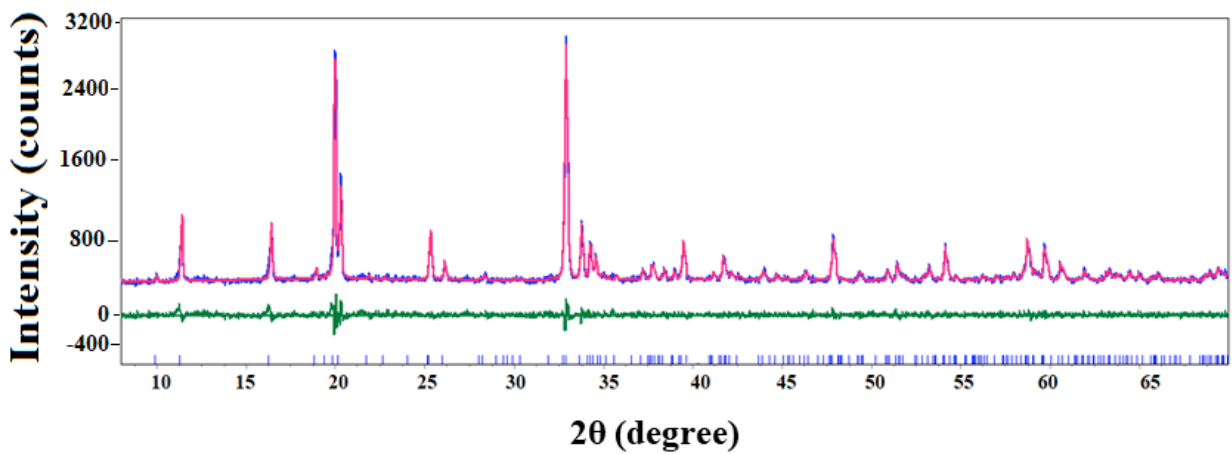

**Figure 1.** Le Bail profile fit of PXRD pattern of $Na_2FeSbO_5$ in orthorhombic space group *Pbna* (No. 60). Observed (blue), calculated (pink), and difference (green) plots are shown, and Bragg reflections are indicated by light blue tick marks.

11.0995(5) Å; and $c$ = 14.7982(7) Å.[46] Our attempts to match the observed PXRD pattern of $Na_2FeSbO_5$ using the structural model of $Ca_2FeCoO_5$ were not successful.

Subsequently, we carried out experiments to grow single crystals. Trial experiments based on high temperature (1200 °C) melting and slow cooling resulted in the formation of $NaFeO_2$ and $Fe_2O_3$. Further explorations were carried out by the utilization of fluxes such as NaCl, $Na_2CO_3$, non-eutectic mixture of $Na_2CO_3$ and $K_2CO_3$, $Na_2SO_4$ as well as an equimolar mixture of $Na_2SO_4$ and $K_2SO_4$. The crystal growth temperature varied from 1000 to 1200 °C. Use of $Na_2SO_4$ as flux led to single crystals of $Na_2FeSbO_5$ of appreciable quality after washing with hexane.

### 3.2. Structure

Single crystal measurements using several crystals, obtained from different batches using $Na_2SO_4$ flux yielded the orthorhombic lattice parameters ($a$ = 15.6991(9) Å; $b$ = 5.3323(4) Å; $c$ = 10.8875(6) Å) with *a* axis being the highest cell dimension. The systematic absences suggested the possible space groups, *Pbna, Pmna, Pmma, Pnnm, Pmm2, Pmc2,*

*Pmn2, Pca2₁* and *P222₁*. SHG measurements performed on the polycrystalline sample of Na$_2$FeSbO$_5$ did not show any significant SHG response with respect to KDP and implied the possibility of the structural solution in a centrosymmetric space group. Out of the various centrosymmetric space groups, the space group *Pbna* (No. 60) was confirmed to be the correct one (Tables 1, 2 and S1).

The Sb1, Sb2, Na1 and Na2 were located at *4c* positions (site symmetry, 2) with the Fe1, Na3 and all the five oxygen atoms (O1 to O5) occupying the general positions (Table 2). The Sb1 and Sb2 atoms form Sb1O$_6$ and Sb2O$_6$ octahedra respectively, which alternate along the b direction by sharing their edges (through O2 and O4) to form zigzag chains (Figure 2). The Fe atoms make FeO$_4$ tetrahedra by sharing their corners (O5) to form zigzag chains along *b*. The zigzag chains of edge-sharing SbO$_6$ octahedra are connected with those of FeO$_4$ tetrahedra by sharing corners (O1 and O3) to form the three-dimensional (3D) framework of Na$_2$FeSbO$_5$, which exhibits the connectivity of -Sb1-O1-Fe-O3-Sb2-O3-Fe-O1-Sb1- along the *c* direction (Figure 2). In addition to the O2 and O4 atoms, the octahedral coordination around Sb1 and Sb2 is completed by the O1 and O3 atoms respectively (Table 3). The sodium atoms, Na1, Na2 and Na3 are located at the pockets of the 3D FeSbO$_5$ framework (Figures 2, 3). In essence, Na$_2$FeSbO$_5$ has the zigzag chains of edge-sharing SbO$_6$ octahedra condensed with those of corner-sharing FeO$_4$ tetrahedra by sharing their corners (Figure 4), with two distinct mirror-related configurations for the zigzag chains of corner-sharing FeO$_4$ tetrahedra (L and R) as found for the brownmillerite structure.[46]

**Table 1.** Crystal data and structure refinement parameters for Na$_2$FeSbO$_5$ based on single crystal X-ray diffraction measurements at 150 K.

| | |
|---|---|
| Formula weight (g/mol) | 303.58 |
| Crystal system | Orthorhombic |
| Space group, Z | *Pbna*, 8 |
| a [Å] | 15.6991(9) |
| b [Å] | 5.3323(4) |
| c [Å] | 10.8875(6) |
| V[Å$^3$] | 911.42 (10) |
| $\rho_{calc}$ [g/cm$^3$] | 4.425 |
| Morphology | Irregular |
| Color | Light brown |
| Dimensions (mm) | 0.09 × 0.07× 0.04 |
| Absorption coefficient | 9.220 mm$^{-1}$ |
| Temperature [K] | 150(2) |
| Wavelength [MoK$_\alpha$] [Å] | 0.71073 |
| Monochromator | Graphite |
| Scan mode | ω scan |
| θ range [°] | 3.743-29.424 |
| hkl range | -19 to 20, -7 to 7, -14 to14 |
| F(000) | 973 |
| R$_{int}$ | 0.1010 |
| R$_{sigma}$ | 0.0656 |
| Refinement | F$^2$ |
| No. of reflections used | 11962 |
| Unique reflections | 1186 |
| Reflections with I ≥ 3σ (I) | 5930 |
| Number of parameters | 59 |
| GoF on F$^2$ | 1.157 |
| R[F$^2$> 2σ(F$^2$)] | 0.0874 |
| wR$_2$ | 0.2386 |
| Largest diff. peak (0.82 Å from Sb1) and hole [e /Å$^3$] | 7.85/-2.915 |

**Table 2.** Atomic coordinates and isotropic equivalent displacement parameters for Na$_2$FeSbO$_5$ based on single crystal X-ray diffraction measurements at 150 K.

| Atom | Wyckoff position | x/a | y/b | z/c | SOF | U$_{eq}$(Å$^2$) |
|---|---|---|---|---|---|---|
| Sb1 | 4c | 0.22241(9) | 0.2500 | 0.5000 | 1.0 | 0.0110(5) |
| Sb2 | 4c | 0.32697(9) | -0.2500 | 0.5000 | 1.0 | 0.0101(5) |
| Fe | 8d | -0.05821(12) | -0.2596(4) | 0.2751(2) | 1.0 | 0.0105(6) |
| Na1 | 4c | 0.0934(7) | -0.2500 | 0.5000 | 1.0 | 0.030(3) |
| Na2 | 4c | 0.4509(6) | 0.2500 | 0.5000 | 1.0 | 0.026(3) |
| Na3 | 8d | 0.3218(5) | 0.3139(14) | 0.7545(6) | 1.0 | 0.0248(17) |
| O1 | *8d* | 0.1376(6) | 0.121(2) | 0.6120(9) | 1.0 | 0.018(2) |
| O2 | *8d* | 0.3156(6) | 0.067(2) | 0.5932(9) | 1.0 | 0.016(2) |
| O3 | *8d* | 0.4170(7) | -0.1342(19) | 0.3857(10) | 1.0 | 0.014(2) |
| O4 | *8d* | 0.2377(6) | -0.0692(19) | 0.4036(9) | 1.0 | 0.013(2) |
| O5 | *8d* | 0.0423(6) | -0.0995(19) | 0.3085(10) | 1.0 | 0.015(2) |

The +5 oxidation state of Sb is supported by bond valence sum (BVS) calculations[47] yielding 5.28 and 5.31 respectively for Sb1 and Sb2. The $FeO_4$ tetrahedra was found to be distorted having four different Fe-O bond lengths varying from 1.83 to 1.91 Å with O-Fe-O angles varying from 107.45° to 123.18° (Table S2, Supporting Information). Here again, the calculated BVS was 2.91, matching well with +3 oxidation state of iron. The charge balancing counter cations (Na1, Na2 and Na3) are having irregular octahedral oxygen coordinations. The Na1, Na2 and Na3 atoms form distorted octahedra; the Na1 and Na2 have three different Na-O bond lengths ranging from 2.37 to 2.67 Å, and the Na3 have six different Na-O bond lengths (2.19 to 2.74 Å) (Table 3).

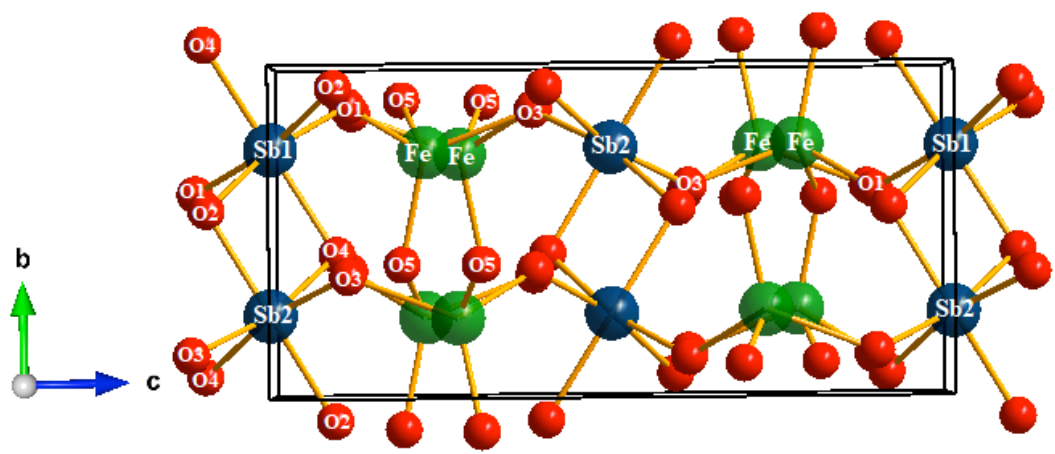

**Figure 2**. Ball and stick representation of the crystal structure highlighting the $FeSbO_5$ framework of $Na_2FeSbO_5$. The $Na^+$ ions are not included for clarity.

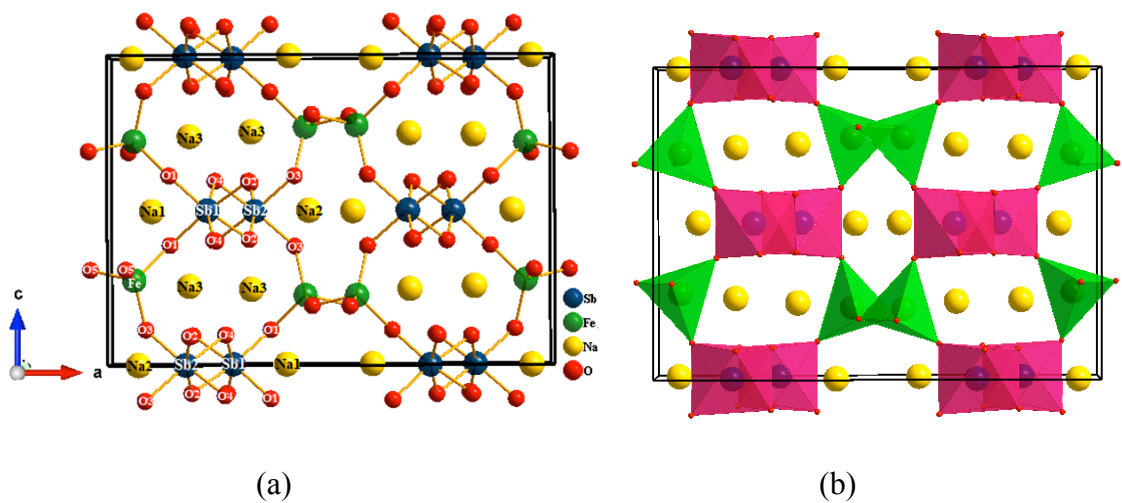

(a) (b)

**Figure 3.** (a) Ball and stick representation of the crystal structure of $Na_2FeSbO_5$ showing the coordination environments of the cations. (b) Corresponding polyhedral representation in the *ac* plane.

**Table 3.** Selected bond distances (in Å) of $Na_2FeSbO_5$ and bond valence sum (BVS) values calculated for the $Sb^{5+}$, $Fe^{3+}$ and $Na^+$ cations.

| Atoms | Bond distance (Å) | BVS |
|---|---|---|
| Sb1–O4 | 2.014(10) × 2 | |
| Sb1–O1 | 1.933(10) × 2 | 5.28 |
| Sb1–O2 | 2.030(10) × 2 | |
| Sb2–O2 | 1.980(11) × 2 | |
| Sb2–O3 | 1.982(10) × 2 | 5.31 |
| Sb2–O4 | 1.999(10) × 2 | |
| Fe–O5 | 1.831(10) | |
| Fe–O5 | 1.865(10) | 2.91 |
| Fe–O1 | 1.901(11) | |
| Fe–O3 | 1.914(11) | |
| Na1–O5 | 2.374(11) × 2 | |
| Na1–O1 | 2.423(11) × 2 | 0.99 |
| Na1–O4 | 2.676(14) × 2 | |
| Na2–O3 | 2.455(11) × 2 | |
| Na2–O3 | 2.497(13) × 2 | 0.92 |
| Na2–O2 | 2.548(13) × 2 | |
| Na3–O2 | 2.196(12) | |
| Na3–O1 | 2.279(12) | |
| Na3–O4 | 2.561(12) | |
| Na3–O5 | 2.491(12) | 1.27 |
| Na3–O4 | 2.283(12) | |
| Na3–O3 | 2.735(12) | |

Finally, the Le Bail PXRD pattern of the bulk crystallites possessing hexagonal morphology (Figure S1, Supporting Information) confirms the formation of single phase; the reflections match well with the orthorhombic space group (Figure 1). Further confirmation was also obtained by comparison between the PXRD pattern of the bulk polycrystalline sample $Na_2FeSbO_5$ with those generated from the single crystal measurements and with those generated from the single crystal structure solution in Figure S2 (Supporting Information).

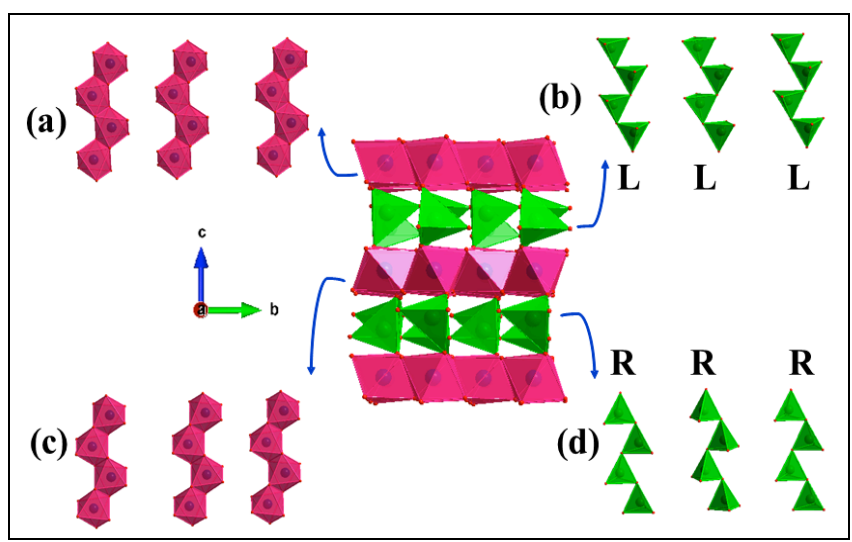

**Figure 4.** Crystal structure of $Na_2FeSbO_5$ in the *bc* plane. (a) Zigzag octahedral chain made up of edge shared $Sb_2O_{10}$ units at z = 0. (b) Zigzag chains of corner-sharing $FeO_4$ terahedra at z = 0.25. (c) Octahedral chains at z = 0.50. (d) Tetrahedral chain at z = 0.75. All the chain orientations are shown in the *ab* plane.

### 3.3. Magnetic susceptibility

The *dc* magnetic susceptibility $\chi = M/B$ measured at different magnetic fields are presented in Fig. 5(a). The clearest feature is a step-like anomaly at around 100 K at $B = 0.1$ T. Below this temperature the magnetic susceptibilities $\chi(T)$, recorded in zero-field-cooled (ZFC) and field-cooled (FC) regimes at low fields, diverge appreciably. This bifurcation tends to close with increasing the strength of the external field and completely disappears at $B = 9$ T. The $\chi(T)$ measured at $B = 1$ T exhibits an additional anomaly at ~ 30 K, the presence of which can be readily recognized by the derivative curve $d\chi/dT(T)$ (see lower inset in Fig. 5(a)). At the same time, there is no sign for a conventional long-range magnetic ordering down to 2 K. As can be seen from Fig. S3 in the Supporting Information, the inverse magnetic susceptibility $1/\chi$ as well as the product $\chi T$ clearly demonstrate the absence of the Curie-Weiss type behavior up to the highest temperature reached in the experiment. Attempts to describe the $\chi(T)$ using the Curie-Weiss law, which crucially depends on the temperature interval used, resulted in unrealistic parameters.

In order to explore the spin dynamic behavior, we have performed *ac* magnetic susceptibility measurements. It is obvious from Fig. 5(b) that there are two smooth but distinct spin-crossover peaks at $T_{f1}$ and $T_{f2}$ in the real $\chi'$ and imaginary $\chi''$ parts of the *ac* magnetic susceptibility, which are both frequency dependent. The values $T_{f1}$ and $T_{f2}$, estimated from the real part $\chi'$ at 0.5 Hz, are approximately 80 and 35 K, respectively. With increasing the frequency, the curves shift to high temperatures. In the investigated frequency range, the magnitude of this shift as measured by the factor $\Delta T_f/T_f\Delta(\log\omega)$ as in canonical spin glasses,[48,49] is 0.04 for $T_{f1}$ and 0.02 for $T_{f2}$. The two critical temperatures differ slightly from those determined from the anomalies in the *dc* $\chi(T)$ measured at $B = 1$ T (Fig.5(a)). The corresponding anomalies on the imaginary part $\chi''$ occur at 73 K and 22 K. Such a decrease in the temperature where the imaginary $\chi''$ peak occurs is characteristic of insulating spin-glasses.[50] The activation energies for the two different relaxation processes can be determined using the Arrhenius law:

$$\omega = \omega_0 \exp[-E_a/k_B T_f] \qquad (1)$$

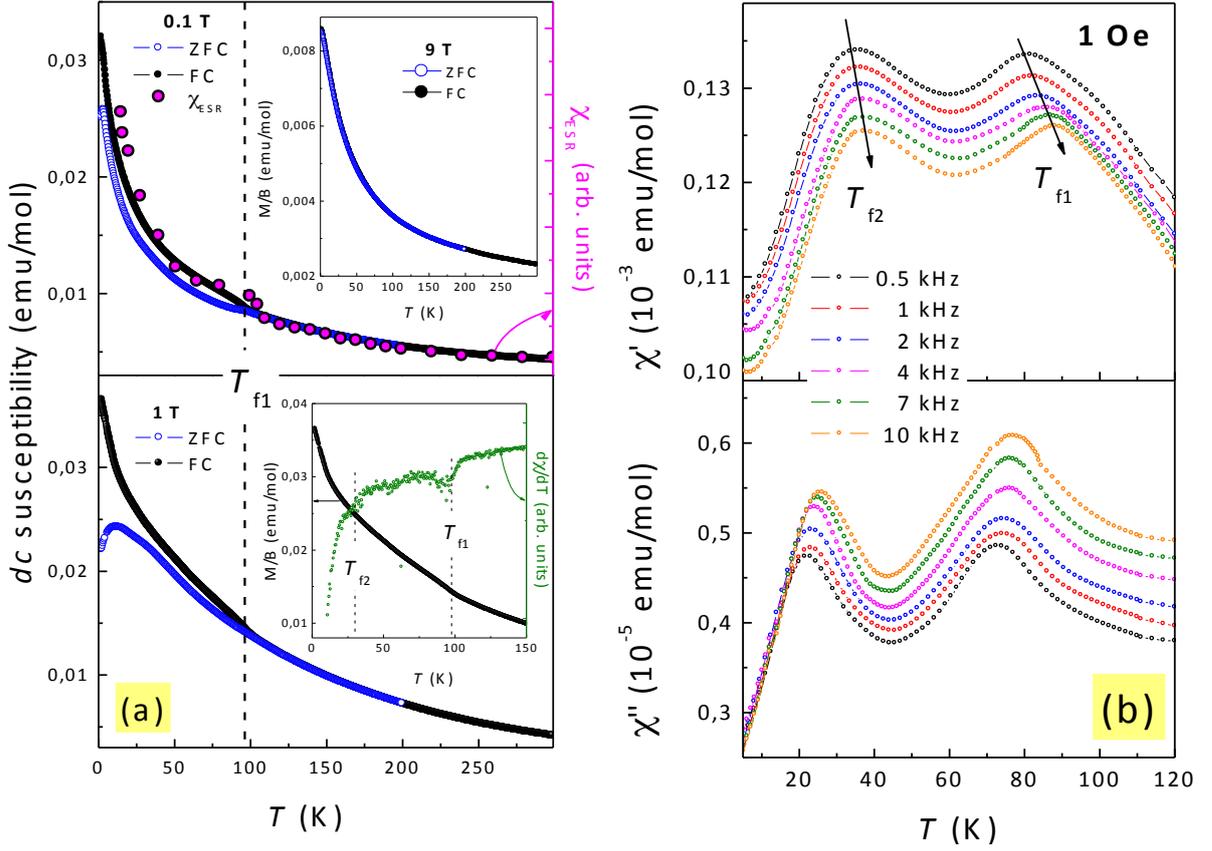

**Figure 5**. Magnetic susceptibility measured for $Na_2FeSbO_5$: (a) Temperature dependence of the *dc* magnetic susceptibility $\chi = M/B$ recorded in the ZFC (blue open symbols) and FC (black filled symbols) regimes at various fields along with dynamic magnetic susceptibility $\chi_{ESR}$ obtained from ESR measurements (magenta filled symbols). (b) Real $\chi'$ and imaginary $\chi''$ parts of the *ac* magnetic susceptibility at various frequencies.

Here $\omega = 2\pi f$ is the driving frequency of our $\chi_{ac}$-measurements, and $T_f$ the peak temperature. Using the Arrhenius plot, $\ln(\omega)$ vs $1/T_f$, we obtain $E_{a1} = 1680 \pm 10$ K for the "high"-temperature freezing, and $E_{a2} = 940 \pm 10$ K for the "low"-temperature freezing (Fig. S4).

It is interesting to note that such a unusual spin dynamics has been observed earlier for several quasi 1D magnetic systems associated with the formation of "partially disordered antiferromagnetic structure", which can be considered as a spin liquid like state. Examples include $KCr_3As_3$,[51] $Sr_{1-x}Ca_xNi_2V_2O_8$,[52] $Ca_3CoRhO_6$[53], $([MnTPP][TCNE])$[54] and $FeMgBO_4$.[55] In a representative system $FeMgBO_4$ consisting of zigzag chains of magnetic ions $Fe^{3+}$, impurities break the magnetic chains. Mössbauer experiments show that the spins of the broken zigzag chains are frozen at low temperature. However, the specific heat measurements show no anomaly, and neutron diffraction measurements on powder samples no Bragg peaks.[55] This indicates the spin glass nature of $FeMgBO_4$.

The frequency dependence of $T_f$ can be described by the critical ''slowing down'' mechanism of spin dynamics,[49,56] which is described by

$$\tau(T_f) = \tau_0 \left( T_f / T_g - 1 \right)^{zv} \quad (2)$$

where $\tau$ is the spin-relaxation time ($\tau \approx 1/f$), $T_g$ the critical temperature for spin-glass ordering at $f \to 0$, $zv$ the dynamical exponent, and $\tau_0$ the characteristic time scale for the spin dynamics. Plots $\log\tau$ vs. $\log[(T_f/T_g)-1]$ are shown in Fig. 6. These plots are best described by the $T_{g1} = 70 \pm 5$ K and $T_{g2} = 30 \pm 5$ K. From the intercept and slope of the fitted straight line, we obtain $\tau_0^1 \sim 10^{-8}$ s, $zv_1 \approx 7$ and $\tau_0^2 \sim 10^{-11}$ s, $zv_2 \approx 10$. The value of $zv$ is in good agreement with those reported experimentally and theoretically for low-dimensional spin-glass magnetic systems.[54,56-59]

**3.4. Magnetization isotherms**

The full magnetization isotherm $M(B)$ at $T = 1.8$ K in external fields from -7 T to 7 T (Fig. 7) has a characteristic feature of spin-glass compounds, namely, an S-shape, implying the presence of a weak ferromagnetic component in the exchange interactions. It also shows a tiny hysteresis with a residual magnetization reaching $M_r \approx 0.002\mu_B$/f.u. (lower inset of Fig. 7). Such a behavior may suggest the freezing of spins. Within this range of the applied magnetic fields, the magnetization isotherm does not display saturation and the magnetic moment is still far below the theoretically expected saturation value for the high spin state Fe$^{3+}$ ($S$ = 5/2) ion: $M_S = gS\mu_B \approx 5$ $\mu_B$. With increasing temperature, the magnetization isotherms $M(B)$ gradually straighten, demonstrating the decrease in the ferromagnetic correlation contribution (upper inset of Fig. 7). No additional magnetic field induced features were detected in the whole temperature range investigated. All the phenomena observed consistently point to a spin-glass state at low temperatures.

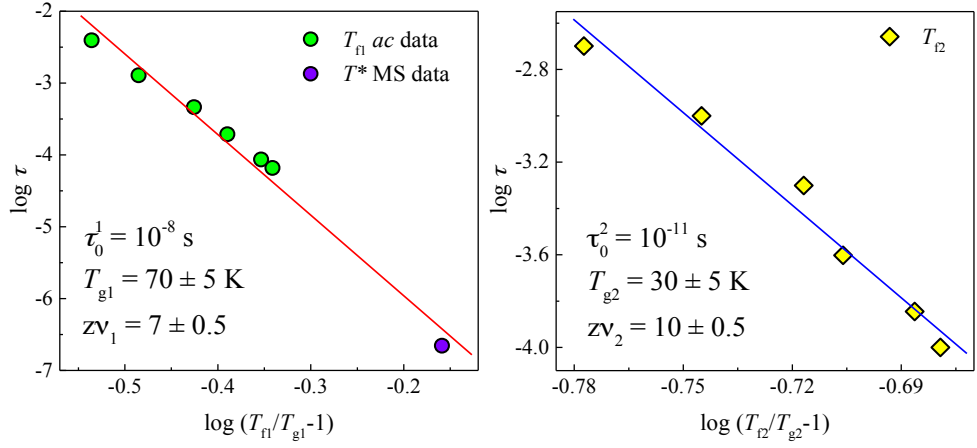

**Figure 6.** Plots of log$\tau$ vs log[($T_f/T_g$)-1] for the two relaxation processes of Na$_2$FeSbO$_5$ at low temperatures The solid lines represent the least-square fit to the experimental data.

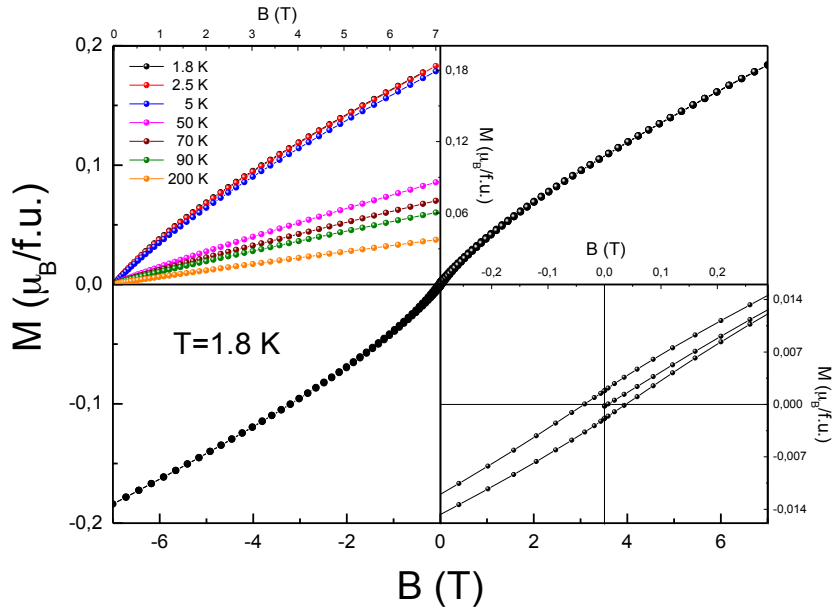

**Figure 7**. The full $M(B)$ isotherm at $T$ = 1.8 K for Na$_2$FeSbO$_5$. Insets: $M(B)$ isotherms at various temperatures and the zoomed-in central part of $M(B)$ at $T$ = 1.8 K.

### 3.5. Specific heat

The temperature dependence of the specific heat $C_p(T)$ for $Na_2FeSbO_5$ at $B$ = 0 T and 9 T is shown in Fig. 8. Over the whole temperature range studied, the specific heat data show no $\lambda$-type anomaly that can indicate a transition to a magnetically ordered state, which is in good agreement with the temperature dependence of magnetic susceptibility. The number of atoms per formula unit in $Na_2FeSbO_5$ is $\nu = 9$, so a classical Dulong-Petit saturation value is expected to be $C = 3R\nu = 224$ J/(mol K), where $R = 8.31$ J/mol K is the gas constant. In low temperature range, the $C_p(T)$ shows a weak broad maximum (Schottky-type anomaly) at $T_{Sch}$ ≈ 5 K, which corresponds most likely to a trace amount of some defects in the $Na_2FeSbO_5$. The application of an external magnetic field does not practically affect the character of the $C_p(T)$ dependence, but slightly shifts the $T_{Sch}$ value to lower temperatures (upper inset in Fig. 8). To get some quantitative estimations, we approximated low-temperature part of specific heat by the sum of the phonon term $C_{ph}$ (in the frame of Debye model) and the magnetic contribution $C_m$ from the Schottky-type anomaly $C_{ph} = \beta T^3 + nC_{Sch}$, where $n$ is the concentration of defects responsible for the Schottky anomaly with energy gap $\Delta$,[60]

$$C_{Sch} = R\left(\frac{\Delta}{T}\right)^2 \frac{e^{(\Delta/T)}}{\left[1+e^{(\Delta/T)}\right]^2} \qquad (1)$$

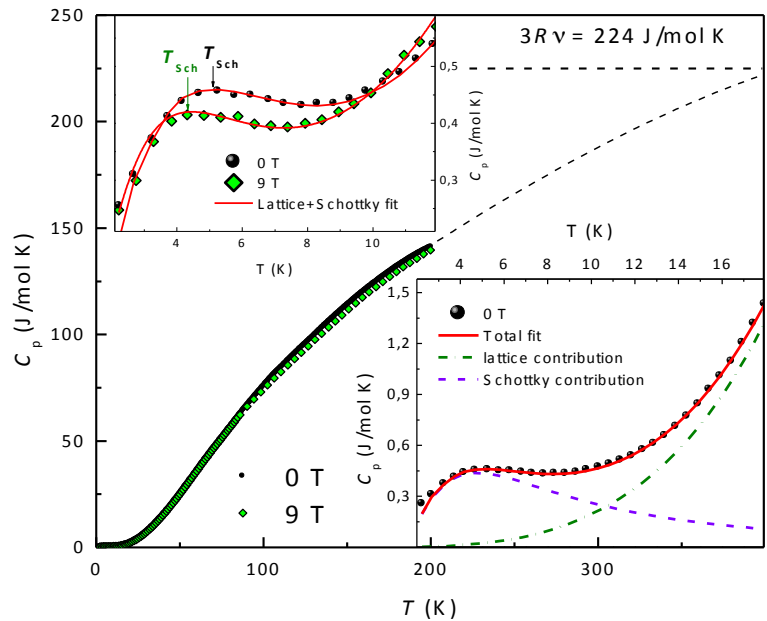

**Figure 8**. The temperature dependence of the specific heat $C_p(T)$ at $B$ = 0 T (black circles) and at $B$ = 9 T (green diamonds) measured for $Na_2FeSbO_5$. The upper inset highlights a Schottky-type anomaly on $C_p(T)$ at low temperatures. The lower inset shows the $C_p(T)$ at $B$ = 0 T in the low temperature range and its approximation, where the green dash-dotted line corresponds to the lattice contribution to the specific heat $C_{ph}$, the violet dashed line the magnetic contribution to the specific heat $C_m$, and the red line their sum.

As can be seen from the lower inset of Fig. 8, the red solid curve gives a reasonable description of the experimental data with the parameters $n ≈ 9.8\%$, $\beta = 1.77×10^{-4}$ J/(mol K$^4$), and $\Delta = 9.9 ± 0.1$ K at zero magnetic field. The external field $B = 9$ T reduces the energy gap down to $\Delta = 8.5 ± 0.1$ K. The estimated energy gap is in reasonable agreement with the position of $T_{Sch} = 0.42\Delta$.[60] Using the parameter $\beta$, the Debye temperature $\Theta_D$ is estimated to be 460 ± 10 K.

### 3.6. Mössbauer spectroscopy

The $^{57}$Fe Mössbauer spectra of $Na_2FeSbO_5$ recorded above 105 K (Fig. 9) can be described as a superposition of two quadrupole doublets, Fe(1) and Fe(2), corresponding to the high-spin ferric ions $Fe^{3+}$ at tetrahedral and octahedral sites, respectively.[61] The second Fe(2) doublet can arise either from the $Fe^{3+}$ ions substituted for the $Sb^{5+}$ ions at $4c$ positions, or from the $Fe^{3+}$ ions located on the surface of the sample grains.

Due to the fact that the contribution of the Fe(2) subspectrum is small (~4%), it is difficult to accurately determine its hyperfine parameters; therefore, this contribution will not discussed further. The parameters of both quadrupole doublets are collected in Table 4 and can be assigned to the high-spin $Fe^{3+}$ ($3d^5$, $S = 5/2$) cations in octahedral and tetrahedral oxygen coordination sites, which are in good agreement with the parameters found for other iron-containing oxides.[62] It is important to note that, according to the Mössbauer spectroscopy data, practically all iron ions (95-96%) are at the tetrahedral sites, consistent with the crystal structure.

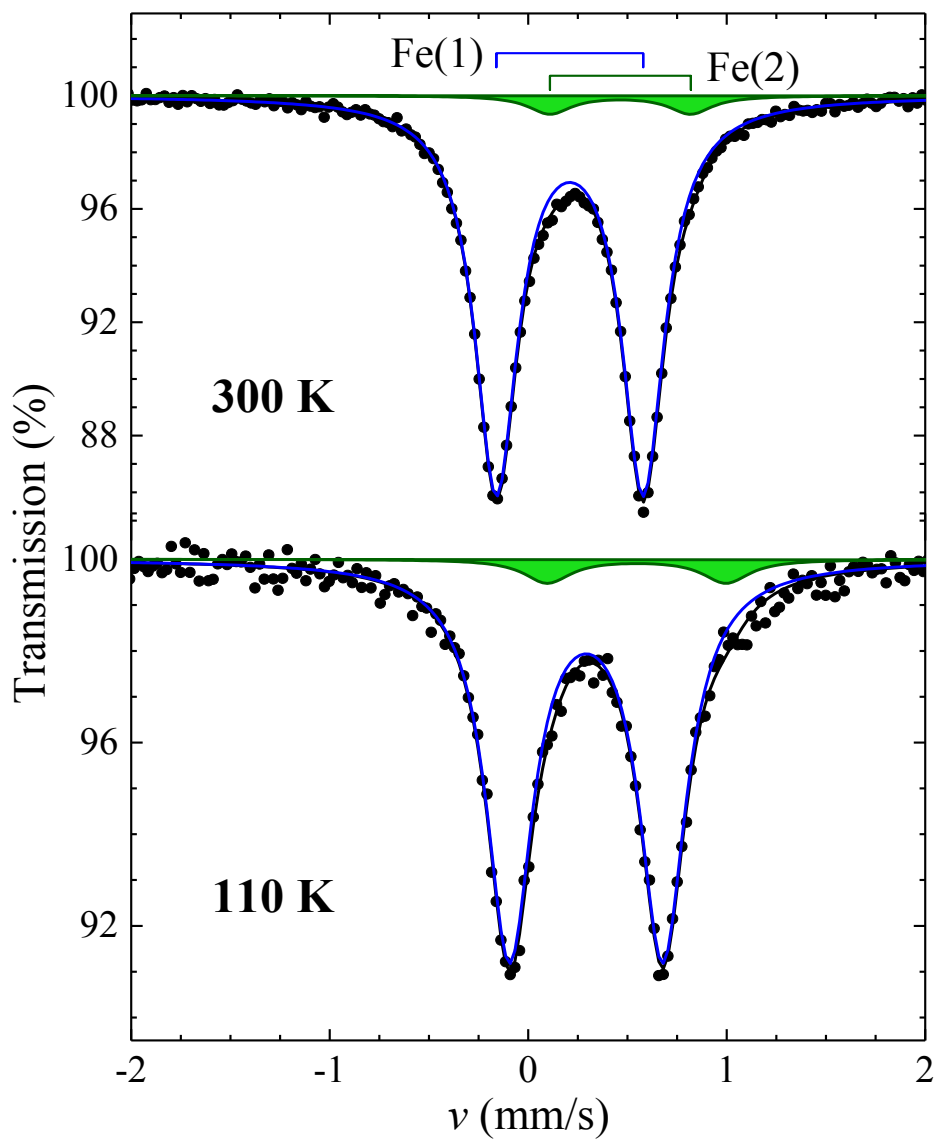

**Figure 9.** $^{57}$Fe Mössbauer spectra of $Na_2FeSbO_5$ recorded in the paramagnetic temperature range above $T^* = 104$ K.

**Table 4**. Hyperfine parameters of the $^{57}$Fe Mössbauer spectra of Na$_2$FeSbO$_5$ at different temperatures.

| $T$ (K) | Subspectrum | $\delta$ (mm/s) | $\Delta$ (mm/s) | $\Gamma$ (mm/s) | $I$ (%) |
|---|---|---|---|---|---|
| 300 | Fe(1) | 0.21(1) | 0.74(1) | 0.26(1) | 95.6(3) |
|  | Fe(2) | 0.46(1) | 0.72(2) | 0.26(1)* | 4.5(3) |
| 110 | Fe(1) | 0.29(1) | 0.77(1) | 0.29(1) | 94.3(9) |
|  | Fe(2) | 0.55(1) | 0.90(4) | 0.29(1)* | 5.7(9) |

$\delta$ is an isomer shift, $\Delta$ is a quadrupole splitting, $\Gamma$ is a linewidth and $I$ is a relative intensity.

*The values were fixed.

Below 105 K, the spectra broaden and show six lines characteristic of magnetic hyperfine interactions (see Fig. S5(a)). A Zeeman structure with broadened components evidences the existence of a distribution of hyperfine magnetic fields $H_{hf}$ at the $^{57}$Fe nuclei, as often observed in spin-glasses.[63] In the first stage of the spectral analysis, we reconstructed the magnetic hyperfine field distribution $p(H_{hf})$ (Fig. S5(b)), assuming a linear correlation between the quadrupolar shift ($\varepsilon_Q$) of the Zeeman components and the value of $H_{hf}$.[43] From the temperature dependences of the mean field $<H_{hf}(T)>$ and dispersion $D_{p(H)}(T) = \{\sum p(H_{hf})(H_{hf} - <H_{hf}>)^2 \delta H_{hf}\}^{1/2}$ (Fig. 10) of the resulting distributions $p(H_{hf})$, we evaluated the temperature $T^* = 104(4)$ K, at which the magnetic hyperfine structure totally vanishes. The temperature $T^*$ cannot be considered as a Néel or Curie temperature because, according to magnetic and thermodynamic measurements, no long-range order is present in Na$_2$FeSbO$_5$. This effect can be associated with the characteristic Mössbauer hyperfine Larmor frequency of ~280 MHz instead a spin freezing temperature ($T_g$).

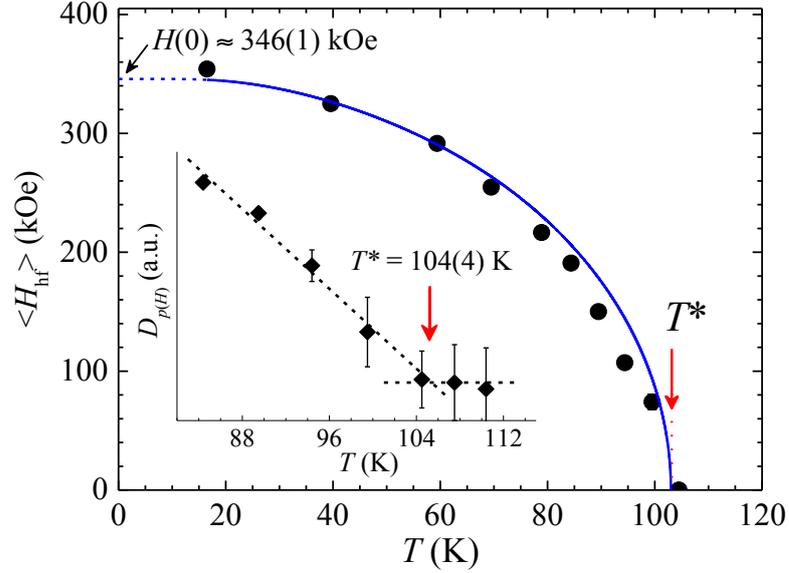

**Figure 10.** Temperature dependence of the average hyperfine field $\langle H_{hf}(T)\rangle$ and dispersion $D_{p(H)}$ (inset) of the distributions $p(H_{hf})$. The solid blue curve corresponds to the fit using the Brillouin function with $S = 5/2$.

The temperature dependence of $\langle H_{hf}(T)\rangle$ was described by Brillouin function (Fig. 10):

$$\langle H_{hf}(T)\rangle = H_{hf}(0) B_S\left[\frac{\sigma(\tau)\cdot S}{\tau\cdot(S+1)}\right], \qquad (3)$$

where $S = 5/2$ for the $Fe^{3+}$ ion, $\sigma = \langle H_{hf}(T)\rangle/H_{hf}(0)$ is the reduced field, $\tau = T/T^*$ is the reduced temperature, and $H_{hf}(0)$ is the "saturation" hyperfine magnetic field. The approximated saturation value of $H_{hf}(0) \approx 346$ kOe is anomalously low for high-spin ferric ions in tetrahedral oxygen environment for which $H_{hf}(0)$ is about 480-520 kOe, corrected for covalence effects.[62] This spin reduction (~ 30 %) cannot be caused by crystal field or covalency effects only and may be attributed to zero-point spin reduction, which has been predicted in quasi-one-dimensional systems to be large.[65]

In the temperature range 40 K $< T < T^*$, the spectra clearly demonstrate relaxation behavior (Fig. 11). We observe an increase in the ratio $\omega_{1,6}/\omega_{3,4}$ of the split between 1st - 6th ($\omega_{1,6}$) and 3rd - 4th ($\omega_{3,4}$) lines, thus evidencing the presence of anisotropic hyperfine field fluctuations.[66] This ratio should not depend on temperature in absence of fluctuation effects. It is important, that such behavior even in the presence of a static magnetically split spectra series cannot occur. These features can be interpreted by persistence of slow collective magnetic fluctuations below $T^*$, which is quite reasonable considering the low-dimensional and frustrated topology of the $Fe^{3+}$ spins in the $Na_2FeSbO_5$ structure.

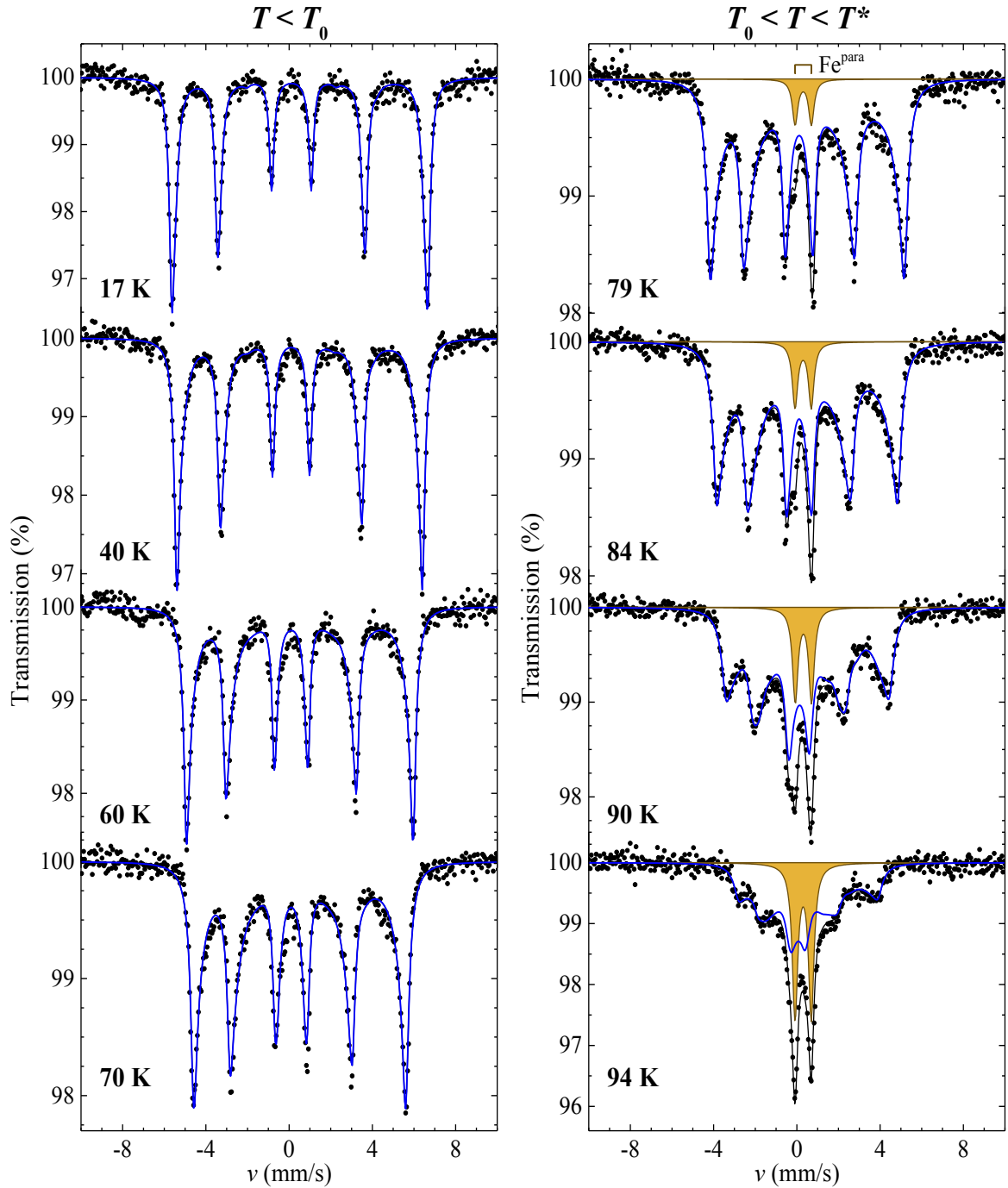

**Figure 11**. $^{57}$Fe Mössbauer spectra of Na$_2$FeSbO$_5$ recorded at $T < T_0$ (left side) where the solid blue lines are the simulations of the experimental spectra using a stochastic relaxation model, and in the range $T_0 < T < T^*$ (right side) where the solid lines are the simulations of the experimental spectra as the superposition of the magnetic (blue line) and paramagnetic (orange area) subspectra.

We used the stochastic model of ionic spin relaxation to simulate the magnetic relaxation spectra. Earlier, this model was used to explain the shape of the $^{57}$Fe Mössbauer spectra of the ferrites demonstrating a geometric frustration and spin-glass-like behavior.[67] The computation procedure can be found in Refs. 67, 68. The $^6S$ state of the $Fe^{3+}$ ion splits into six levels driven by the Zeeman interaction with the Weiss magnetic field $H_W$; the ratio of the thermal population of the successive Zeeman levels is given by $s = exp(-2\mu_B H_W/k_B T)$. The thermal average $<S_z(T)>$ of the $Fe^{3+}$ ion is given by

$$\langle S_z(T) \rangle = \frac{2.5 + 1.5s + 0.5s^2 - 0.5s^3 - 1.5s^4 - 2.5s^5}{1 + s + s^2 + s^3 + s^4 + s^5} \qquad (4)$$

The rate of flipping between the ionic levels, $\Omega_e$, is related to the relaxation time $\tau$ by the relation $\tau = [7(1 + s)\Omega_e]^{-1}$. All quantities appearing in the formula of the Mössbauer lineshape, except $s$ (or $\tau$) and $\Omega_e$, are obtained by fitting the spectrum at 17 K. The lineshape of the spectrum depends not only on the "static" hyperfine parameters ($\delta$, $\varepsilon_Q$, $\Gamma$), but also on the "dynamical" parameters: $s$, $\Omega_e$, and $H_{hf}$ (the saturated hyperfine field when $s \rightarrow 0$). At higher temperatures, only $s$ and $\Omega_e$ are treated as variable parameters. A satisfactory description of the spectra at all temperatures (Fig. 11) was achieved only assuming that there is a continuous distribution of averaged population $s = exp(-\Delta/kT)$, where $\Delta$ is an energy gap between two neighboring relaxation levels. Fig. 12 represents the thermal variation of the selected distributions $p(\Delta)$ reconstructed by the fits of the spectra shown in Fig. 11. The physical origin of the observed $\Delta$ gap distributions can be the anisotropy of exchange interactions as often observed in spin-glasses systems.[69-71]

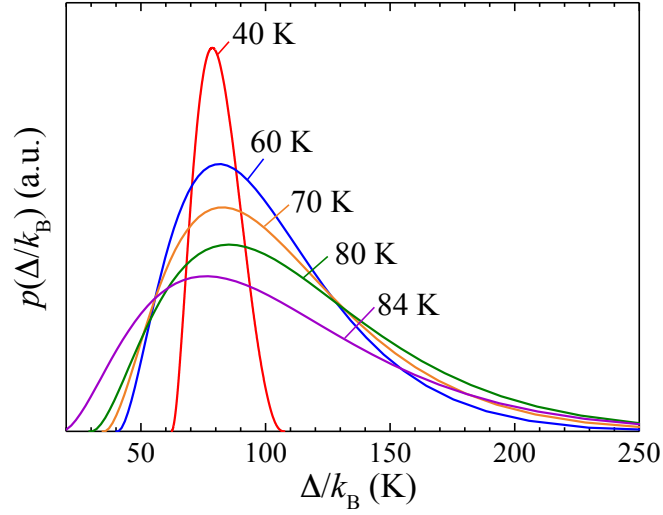

**Figure 12**. Temperature dependence of the distribution $p(\Delta/k_B)$ of the energy gap ($\Delta/k_B$) between two adjacent relaxation levels in the stochastic relaxation model (*see text*). The distribution at 40 K was scaled down by factor 2.5 for the convenience of perception.

The observed thermal variation of $\tau^{-1}(T)$ (see the inset in Fig. 13) can be fitted to the sum of a temperature independent spin-spin term $(\tau^{-1})_0$ and an exponential term associated with a two-phonon process through the excited states, the *Orbach* process[72]: $\tau^{-1}(T) = (\tau^{-1})_0 + B \cdot exp(-U/k_B T)$, with the energy of an excited crystal field level $U$. The fit yields: $(\tau^{-1})_0 = 7.3(2) \times 10^7$ s$^{-1}$, $B \approx 4.5 \times 10^9$ s$^{-1}$ and $U = 41(9)$ meV. Fig. 13 represents the temperature dependence of $<S_z(T)>$ in accordance to Eq.(4). This dependence agrees qualitatively with the thermal variation of the mean field $<H_{hf}(T)> \propto <S_z(T)>$ (Fig. 10) obtained from the $p(H_{hf})$ distribution analysis. The observed thermal variation of $\Omega_e(T)$ (insert in Fig. 13) is fitted as a constant value $8.89(3) \cdot 10^6$ s$^{-1}$, thus indicating that the only spin-spin relaxation occurred in the Na$_2$FeSbO$_5$ lattice.

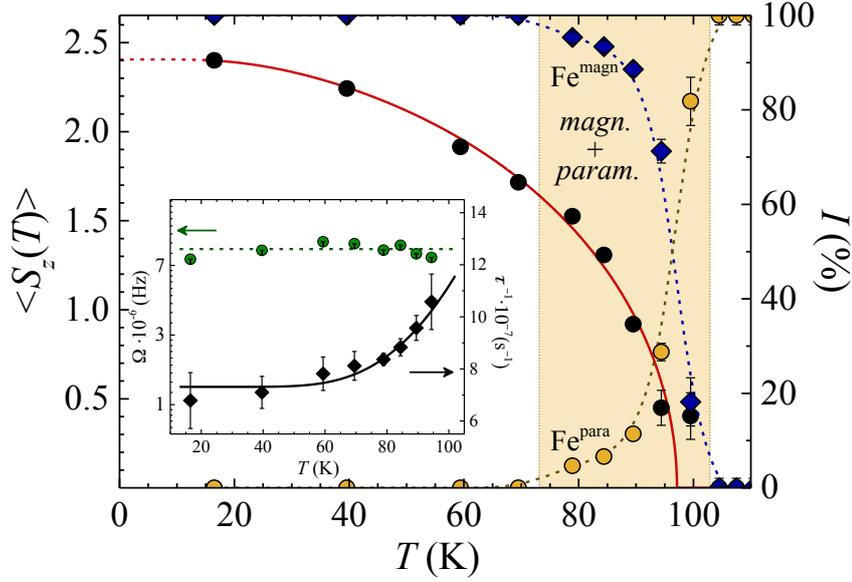

**Figure 13.** Temperature dependence of $<S_z(T)>$ of the $Fe^{3+}$ ions, determined from the values of the thermal population ($s$) of the successive Zeeman levels where the red solid line corresponds to fit using the Brillouin function with $S = 5/2$. Inset: the rate ($\Omega_e$) of flipping between the ionic levels and inverse value of the relaxation time ($\tau$). The solid line corresponds to fit according to: $(\tau^{-1})_0 + B \cdot exp(-U/k_BT)$ (see text).

The $T^*$ value found from Mössbauer spectra appears to be noticeably higher (by ~20%) than that ($T_{f1} \approx 80$ K) obtained from the measurements of magnetic *ac* susceptibility, but agrees well with anomaly in *dc* $\chi(T)$ (Fig. 5). A generic feature of spin-glasses is the critical low (Eq. 2) expected for the characteristic frequency ($\omega = 2\pi f$) of the measurement. Fig. 6 shows the dependence of the frequency $\omega$ on a log scale with the temperature calculated using $T_{g1}(0) = 70(5)$ K. We included the "freezing" temperature $T^* = 104(4)$ K measured by the present $^{57}Fe$ Mössbauer study in order to sweep a larger frequency range. A characteristic frequency $\Omega_e \approx 8.89(3) \cdot 10^6$ s$^{-1}$ deduced from analysis of relaxation spectra (Fig. 11) was used. The critical scaling is seen to be satisfactorily obeyed over 10 orders of magnitude for $\omega$, with reasonable physical values $\omega_0 = 10^8$ Hz and $z\nu = 7.0(5)$.

It is clearly seen that as the $T^*$ temperature is approached from below (starting with $T_0 \approx$ 79 K), a paramagnetic component ($Fe^{par}$) coexisted with the magnetic spectrum ($Fe^{mag}$), continuous to show the presence of ionic spin relaxation, appears (Fig. 11) and sharply grows in intensity ($I$) with temperature (Fig. 13). The transformation of the magnetic spectrum into a paramagnetic one requires that the magnetic sublattice splits into superparamagnetic clusters with more rapid than the nuclear Larmor precession period $\tau_L$ ($\sim 10^{-8}$ s)[73,74] fluctuations of the magnetization. Therefore, the spin freezing in the $Na_2FeSbO_5$ lattice occurs via cluster formation. The observed rapid development of a paramagnetic doublet $Fe^{par}$ is characteristic for a gradual decrease of short range ordering up to temperature $T^*$ where all clusters are successively deblocked. We speculate that the formation of "clusters" reflecting the spin glass state of $Na_2FeSbO_5$ may be described by a random coupling of strongly correlated quasi-one-dimensional Fe-Fe chain segments due to frustration of various magnetic interactions.

### 3.7. ESR spectroscopy

The ESR spectra of $Na_2FeSbO_5$ powder samples measured at various temperatures are presented in Fig. 14. The ESR line shape has a rather complex character and changes radically with the temperature variation. At high temperatures the Lorentzian shape line with a g-factor $g \approx 2$, typical for high-spin state $Fe^{3+}$ ions in tetrahedral coordination, is observed. When temperature decreases the strong distortion of the ESR spectrum is observed. The experimental spectra are relatively broad, hence two circular components of the exciting linearly polarized microwave field on both sides of $B = 0$ should be included to standard Lorenzian profile for proper fitting of the line shape [75]:

$$\frac{dP}{dB} \propto \frac{d}{dB}\left[\frac{\Delta B}{\Delta B^2 + (B - B_r)^2} + \frac{\Delta B}{\Delta B^2 + (B + B_r)^2}\right] \quad (5)$$

where $P$ is the power absorbed in the ESR experiment, $B$ the magnetic field, $B_r$ the resonance field, and $\Delta B$ the line width. A satisfactory description can be achieved by using one, two and three Lorentzian lines in the temperature ranges of 6 – 30 K, 220 – 450 K and 40 – 220 K, respectively. The results of ESR line shape fitting are shown by the solid curves in Fig. 14, which shows a good agreement between the fitted curves and the experimental data. Representative examples of the ESR spectrum decomposition are given in Fig. 15. The main contributions to the absorption come from the two main resonance modes, the $L_1$ and $L_2$ lines (the green and blue curves in Fig. 15, respectively). Moreover, the dominating contribution brings $L_1$ mode over the whole investigated temperature range, which allows us to assign the $L_1$ mode to the signal from the zigzag chains of corner-sharing FeO$_4$ tetrahedra, and the $L_2$ mode to fragments of these chains. The latter manifest themselves as separate paramagnetic subsystem.

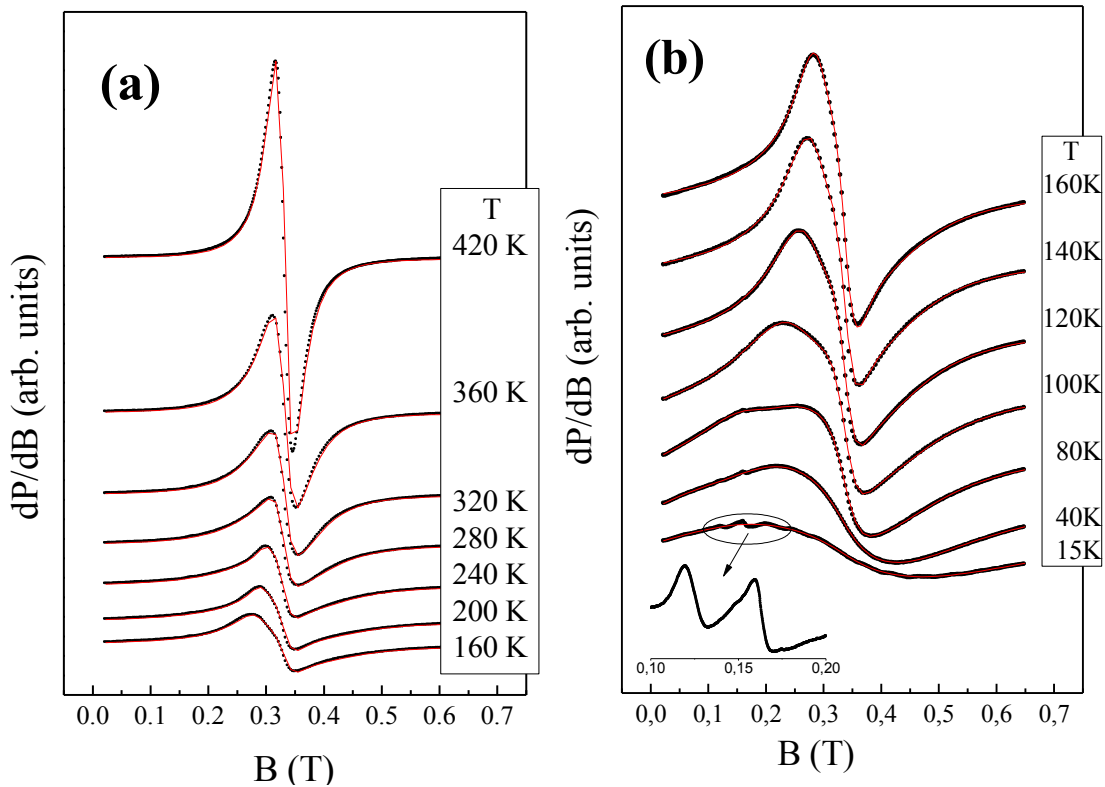

**Figure 14**. Temperature dependence of the first derivative absorption line for Na$_2$FeSbO$_5$. The black circles are experimental data, the lines are results of fitting by the sum of two ($T > $ 225 K, (a)) or three ($T < $ 225 K, (b)) Lorentzians, where each of Lorentzian is given by eq.5.

For both $L_1$ and $L_2$ lines in whole temperature range, the position of the resonance field $B_r$ practically does not change. This indicates the absence of a long-range magnetic ordering down to 6 K. Thus, the ESR data agree well with the static magnetization and specific heat data.

In addition, a careful analysis of the ESR spectra reveals the appearance of the another $L_3$ line, whose amplitude also rapidly grows with decreasing $T$, and the resonant field is strongly shifted towards weaker fields, in contrast to the resonance fields of $L_1$ and $L_2$ lines that remain practically unchanged over the entire temperature range. The strong shift of the resonance field for the $L_3$ line indicates the presence of the internal magnetic field, which changes the resonance conditions. It is usually characteristic of a magnetic ordering. In this case, however, it is obvious that this ordering does not capture the whole volume of a sample and most probably responds to the formation of limited ordered areas (linear clusters).

At $T < 20$ K, the main $L_1$ line dominates the spectra, while the $L_2$ and $L_3$ lines practically disappear: the position of the resonance for $L_3$ obviously falls out of the investigated magnetic field range, while the presence of $L_2$ becomes hidden in the background of more intensive $L_1$ line due to the strong broadening of the $L_1$ and $L_2$ lines. Additionally, at lowest temperatures one can see two weak satellite signals at $g_4 \approx 4.1$ and $g_5 \approx 5.6$ (see inset in Fig. 14b) on the background of the main $L_1$ mode. Similar ESR response has been observed earlier in other $Fe^{3+}$ containing oxides[76] and signal at $g \approx 4.2$ has been related to the median Kramers doublet from Fe sites, which are located in either tetrahedral or octahedral distorted coordination. This interpretation has been supported by Loveridge and Parke[77], and a similar behaviour has been observed experimentally for $Na_4FeSbO_6$.[17] Following by this interpretation, we assume that the major signals at $g_1 \approx 2$ and $g_5 \approx 5.6$ are typical for $Fe^{3+}$ ions in tetrahedral oxygen coordination, while the signal at $g_4 \approx 4.2$ might be associated with $Fe^{3+}$ ions in rhombohedrally distorted environments.

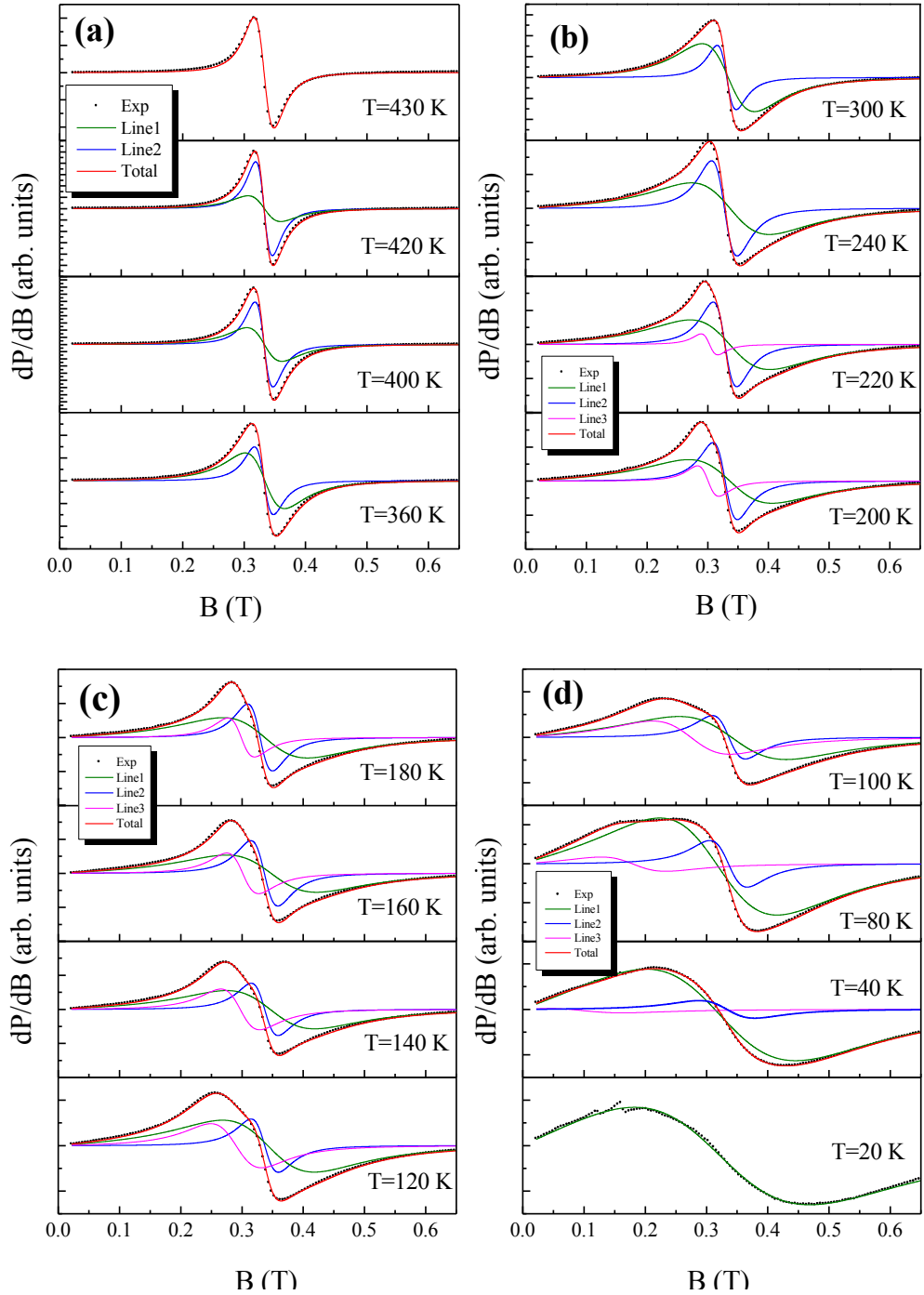

**Figure 15**. ESR spectra of $Na_2FeSbO_5$ at different temperatures. The black circles represent the experimental data ((a): 360 < T < 430 K, (b): 200 < T < 300 K, (c): 120 < T < 180 K, (d): 20 < T < 100 K), and the curves the results of the fitting with each Lorentzian given by Eq.(5). The green, blue and magenta curves represent the resolved components of the ESR spectra, and the red solid line the sum of the components.

The values of the linewidth $\Delta B$ and effective g-factor derived from the fitting analyses for the ESR spectra of $Na_2FeSbO_5$ are collected in Figs. 16-17. In the temperature range of $T > 100$ K, the $L_1$ and $L_2$ curves are characterized by an isotropic temperature-independent effective g-factor $g = 2.00 \pm 0.05$, while the g-factor of the $L_3$ curve is very close to the value $g_3 = 2.2$ (Fig. 16). With decreasing temperature, the temperature dependence of $g(T)$ becomes noticeably different; at $T^* = 100(5)$ K, the g-factor of the $L_1$ curve exhibits a small jump reaching $g_1 = 2.1$, the g-factor of the $L_2$ curve stays almost unchanged, and that of the $L_3$ curve sharply increases. Such behaviors of $g(T)$ are in qualitative agreement with the presence of the main anomaly on the *dc*/*ac* magnetic susceptibility and Mössbauer data, below which spin-freezing ordering occurs.

The temperature dependences of the ESR linewidth for the $L_1$, $L_2$ and $L_3$ curves are shown in Fig. 17. The linewidth $\Delta B_1$ monotonically increases over the entire $T$-range, except

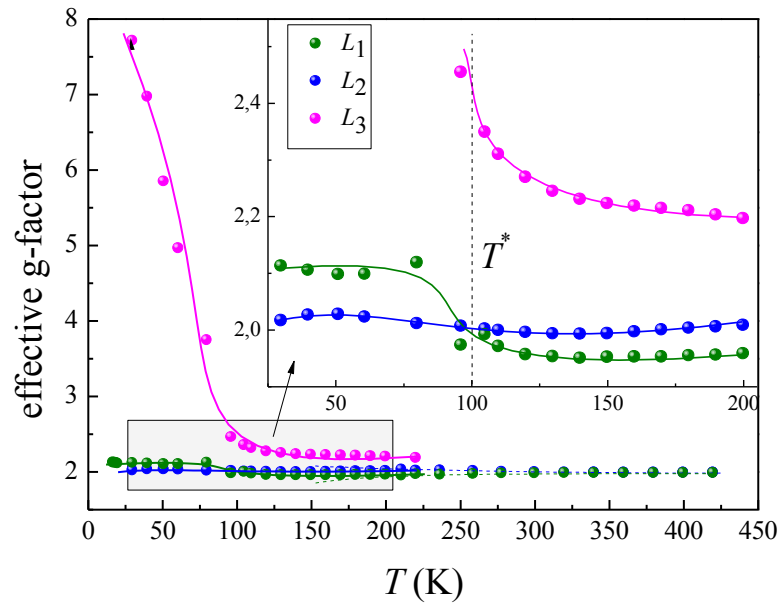

**Figure 16**. The temperature dependence of the effective g-factor for $Na_2FeSbO_5$. The green, blue and magenta circles are obtained from the three Lorentzian fits for $L_1$, $L_2$ and $L_3$, respectively. The inset shows a zoomed-in view in the low temperature range. The dashed line indicates the critical point of the observed anomaly

for 100 K > $T$ > 210 K, where a wide step-like transition to another dynamic regime is observed. The linewidths $\Delta B_2$ and $\Delta B_3$ remain almost temperature-independent down to ~210K and ~100K, respectively, and then increase progressively. Such a critical line broadening for all components is indicative of the development of strong spin-spin correlations and slowing down of spins at low temperatures.[78,79] The same line broadening over the wide temperature range was reported for many other antiferromagnetic compounds, including classical and low-dimensional ones as well as for spin-glass materials.[80-85]

The integral ESR intensity, $\chi_{ESR}$, which is known to be proportional to the concentration of paramagnetic centres, estimated by the double integration was found to agree well with the FC static magnetic susceptibility $\chi(T)$ and also exhibits a step-like anomaly at $T^* \approx 100$ K (Fig. 5(a)).

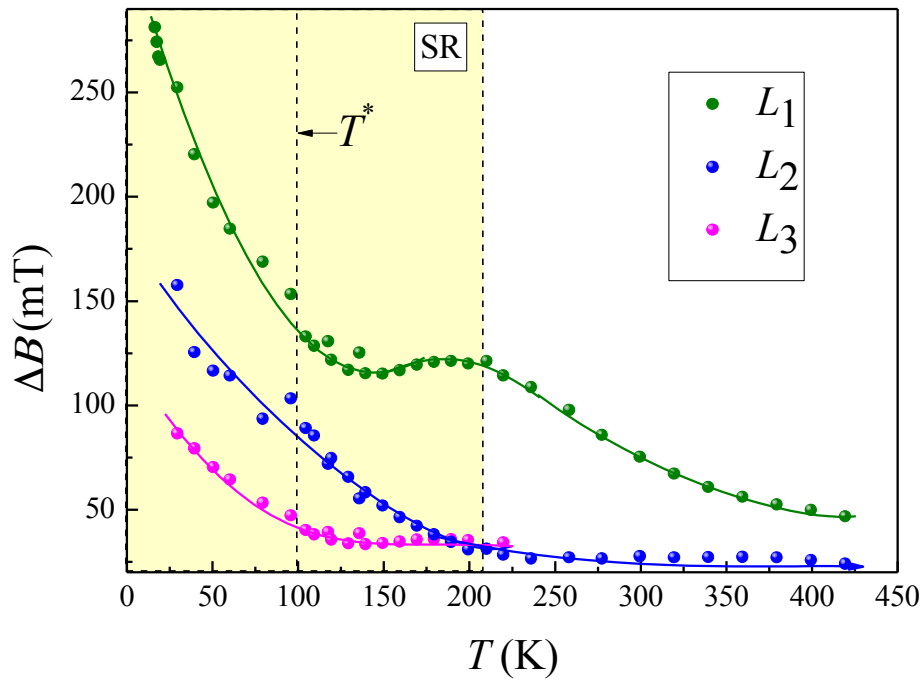

**Figure 17.** The temperature dependence of the ESR linewidth for $Na_2FeSbO_5$. The green, blue and magenta circles are obtained from the three Lorentzian fits for $L_1$, $L_2$ and $L_3$, respectively. The dashed lines indicate the critical temperatures, where anomalies are observed. Yellow area denoted as SR corresponds to the region with the short range fluctuations.

Obviously, the temperature dependence of the ESR parameters implies the significant role of the short-range magnetic correlations at temperatures below 210 K and confirms a transition to slow dynamic regime, i.e., spin-glass state at low temperature.

### 3.8. Density functional analysis of magnetic structure

To explain the observed magnetic properties of the Na$_2$FeSbO$_5$, we determine its spin exchange interactions by carrying out energy-mapping analysis based on DFT calculations.[86-88] The chains of corner-sharing FeO$_4$ tetrahedra lie in the *bc*-plane as depicted in Fig. 18, where the red, cyan, green and yellow circles represent the Fe, O(1), O(3) and O(5) atoms, respectively. As shown in Fig. 19a, the spin exchange paths of interest are the nearest-neighbour (NN) intrachain spin exchange $J_1$ as well as the NN interchain exchanges paths $J_2$ and $J_3$. The structural parameters associated with the exchange paths $J_1 - J_3$ are summarized in Table 5. To determine the values of $J_1 - J_3$, we carry out spin-polarized DFT calculations using a (2a, b, c) supercell for Na$_2$FeSbO$_5$ for the four ordered spin states defined in Fig. 19 by employing the projected augmented wave method encoded in the Vienna Ab Initio Simulation Package (VASP)[89-91] with the generalized gradient approximation of Perdew, Burke and Ernzerhof[92] for the exchange-correlation functionals with a plane wave cutoff energy of 450 eV, a set of 2×4×8 k-points, and a threshold 10$^{-6}$ eV for energy convergence. The DFT plus on-site repulsion U (DFT+U) method[93] was employed at $U^{eff} = U - J = 3$ and 4 eV to describe the electron correlation in the Fe 3*d* states. Given the spin Hamiltonian,

$$H_{spin} = \sum_{i<j} J_{ij} \vec{S}_i \cdot \vec{S}_j \qquad (6)$$

where $\vec{S}_i$ and $\vec{S}_j$ are the spins at magnetic ion sites *i* and *j*, respectively, with the spin exchange constant $J_{ij} = J_1$, $J_2$ or $J_3$. Then, the total spin exchange energies of the four ordered spin states are given by

$$E_{FM} = (-16J_1 - 8J_2 - 8J_3)(N^2/4)$$

$$E_{AF1} = (16J_1 - 8J_2 - 8J_3)(N^2/4)$$

$$E_{AF2} = (16J_1 + 8J_2 - 8J_3)(N^2/4)$$

$$E_{AF3} = (16J_1 - 8J_2 + 8J_3)(N^2/4)$$

per supercell (2a, b, c), where $N = 5$, i.e., the number of unpaired spins at each $Fe^{3+}$ ion site. Thus, by mapping the relative energies of these four states determined by the spin exchange parameters onto the corresponding energies determined by the DFT+U calculations, we obtain the values of $J_1 - J_3$ summarized in Table 6. The spin exchanges $J_1 - J_3$ are all antiferromagnetic (AFM). The intrachain spin exchange $J_1$ is much stronger than the two interchain spin exchanges $J_2$ and $J_3$. Magnetically, $Na_2FeSbO_5$ is best described as a 1D chain system forming the AFM chains made up of the spin exchanges $J_1$ along the *b*-direction. This agrees with the magnetic properties of $Na_2FeSbO_5$ discussed in the previous sections.

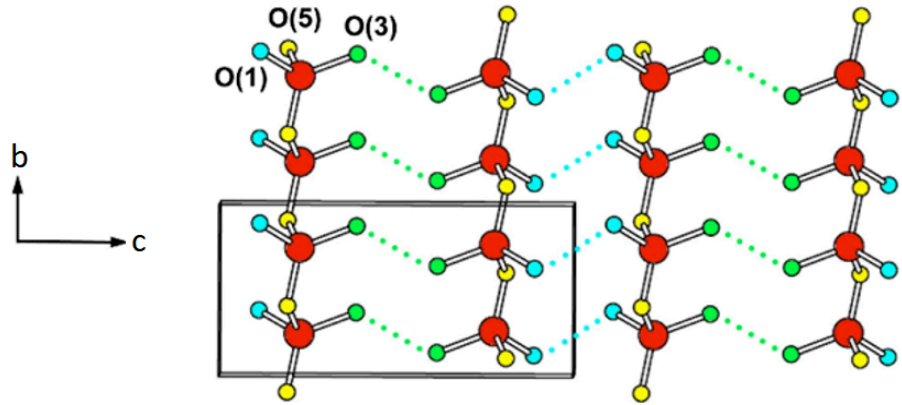

**Figure 18.** Chains of corner-sharing $FeO_4$ tetrahedra in the *bc*-plane, where the red, cyan, green and yellow circles represent the Fe, O(1), O(3) and O(5) atoms, respectively.

**Table 5.** Geometrical parameters associated with the spin exchange paths $J_1$, $J_2$ and $J_3$ of $Na_2FeSbO_5$

|  | Fe…Fe (Å) | O…O (Å) | ∠Fe-O…O (°) | ∠O…O-Fe (°) |
|---|---|---|---|---|
| $J_1$ | 3.233 | | | |
| $J_2$ | 4.898 | 2.800 | 111.9 | 111.9 |
| $J_3$ | 5.991 | 2.778 | 131.6 | 131.6 |

**Table 6.** Spin exchange parameters (in $k_B$K) of $Na_2FeSbO_5$ obtained from GGA+U calculations

|  | $U^{eff}$ = 3 eV | $U^{eff}$ = 4 eV |
|---|---|---|
| $J_1$ | 215.2 | 181.1 |
| $J_2$ | 3.23 | 2.65 |
| $J_3$ | 4.47 | 3.81 |

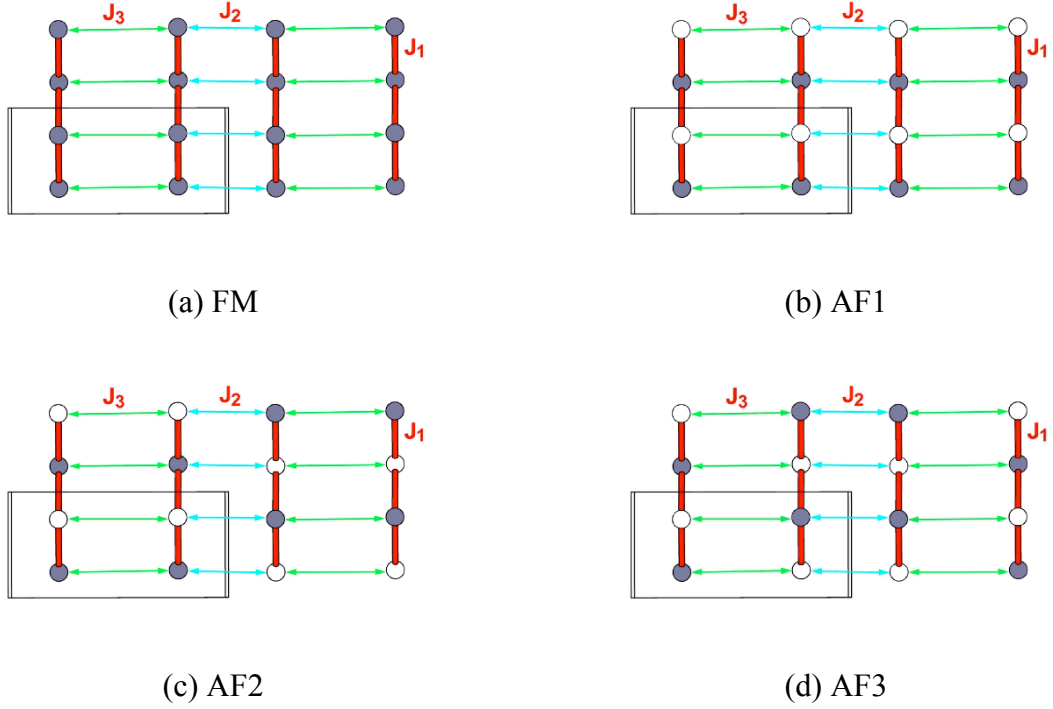

(a) FM  (b) AF1

(c) AF2  (d) AF3

**Figure 19**. Ordered spin arrangements of the FM (a), AF1 (b), AF2 (c) and AF3 (d) states used to extract the values of $J_1 - J_3$. The grey and white circles indicate the spin up and down sites of $Fe^{3+}$ ions.

Using interface provided by the JASS code[94] we simulated temperature dependence of magnetic susceptibility using quantum Heisenberg model for weakly coupled antiferromagnetic spin 5/2 chains. We used quantum Monte-Carlo (QMC) method as realized by the Loop algorithm in the ALPS package[95] to simulate 8x4x4 cell. The exchange parameters obtained in the GGA+U calculations for $U_{eff}$ = 4 eV were used. Number of sweeps was set up to be equal $10^6$. Results are presented in Fig. 20, which shows that the considered spin model does not order until very lowest temperatures used in the calculations (5 K). There is a broad susceptibility maximum reflecting presence of the short-range spin correlations very typical for low-dimensional spin systems, weakly coupled Heisenberg chains in particular.[96,97] These correlations are seen in Mössbauer and ESR spectra. On another hand, temperature behavior of the calculated susceptibility is very different from what is seen in the experiment, see Fig. 5. Theory and experiment can be reconciled if we take into account fragmentation of the chains, as suggested by the ESR. Due to segmentation, the system should rather be considered not as a network of the chains, but as a set of weakly coupled chain fragments (of different lengths), which obviously results in the Curie-like behavior at low temperatures seen in the experiment. At higher temperature these tails subside, but there is still a contribution from those chains, which remain unbroken. Because of large intra-chain exchange coupling the correlation maximum for these chains lies very high in temperature – at $T_{max} \approx 850$ K, much higher than the experimentally investigated temperature range in the present work. This explains why the observed magnetic susceptibility does not obey the Curie-Weiss law up to 400 K.

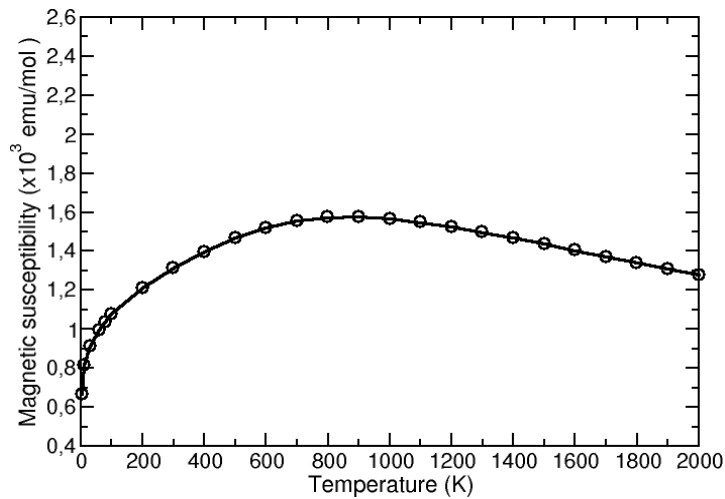

**Figure 20**. Calculated temperature dependence of the magnetic susceptibility of weakly coupled spin 5/2 chains with intra-chain ($J_{intra}$ = 181 K) and inter-chain ($J_{intra}$ = 3 K) exchanges.

## 4. CONCLUDING REMARKS

Single crystals of $Na_2FeSbO_5$ were obtained by using sodium sulphate flux. In this quasi-one-dimensional oxide, zigzag chains of edge-sharing $SbO_6$ octahedra are condensed with chains of corner-sharing $FeO_4$ tetrahedra. The $^{57}Fe$ Mössbauer spectra indicate that all high-spin $Fe^{3+}$ ions occupy equivalent sites in the distorted tetrahedral oxygen coordination. The magnetization, specific heat and ESR data indicate the absence of a long-range magnetic order down to 2 K and extended temperature range of the short-range magnetic correlations. There is neither saturation nor magnetic field induced features on the magnetization isotherms in the whole investigated temperature range. At the same time overall bulk magnetic studies demonstrated a spin-glass-type behavior at low temperatures, which is directly confirmed by *ac* magnetic susceptibility, which manifested a unique two-step freezing at $T_{f1} \sim 80$ K and $T_{f2} \sim 35$ K. Low temperature Mössbauer spectra also reveal a spin-glass behavior below temperature $T^* \approx 104$ K. The theoretical fits of the line shapes in accordance with the stochastic model of ionic spin relaxation show agree well with the experimental MS data. The appearance of a spin-glass transition is supported by frequency dependence of the freezing temperature over a wide frequency range. The spectra just below $T^*$ in the range $T_{f1} < T < T^*$ demonstrate an appearance of quadrupole doublet coexisting with the magnetic subspectrum and sharply growing in intensity with temperature. Such behaviour demonstrates that the spin freezing in the $Na_2FeSbO_5$ lattice occurs via cluster formation. The ESR data yield a complicated picture concerning the coexistence of different magnetic sublattices, which tentatively correspond to the signals from the ordered and disordered $FeO_4$ tetrahedral 1D spin chain clusters. In full agreement with Mössbauer spectra data, both the static magnetic susceptibility and the integral ESR intensity reveal a clear anomaly at $T^* \sim 100$ K. The spin exchanges determined by the mapping analysis based on DFT calculations show a strong 1D AFM character with the main intrachain exchange of ~200 K and small interchain interaction of ~3 K. Our results seem to be related to the fluctuations enhanced by the weakened interchain coupling leading to fractionalization of linear chain clusters and freezing the system into spin-cluster ground state.

## ASSOCIATED CONTENT

**Supporting Information**

The Supporting Information is available free of charge on the ACS Publications website. It contents the figures of FESEM, EDX, comparative PXRD patterns, *dc* magnetic susceptibility $\chi$ and its inverse value $1/\chi$, Arrhenius plots $\ln(\omega)$ vs $1/T$ and $^{57}$Fe Mössbauer spectra at different temperatures; the tables of bond angles and anisotropic displacement parameters.

**Accession codes**

CCDC -1858526 contains the supplementary crystallographic data for this paper. These data can be obtained free of charge via www.ccdc.cam.ac.uk/data_request/cif, or by emailing data_request@ccdc.cam.ac.uk, or by contacting The Cambridge Crystallographic Data Centre, 12 Union Road, Cambridge CB2 1EZ, UK; fax: +44 1223 336033.

## ACKNOWLEDGMENTS

Financial support received from DST- SERB (EMR/2016/006762) is gratefully acknowledged. SU and AS thank University Science and Instrumentation Centre (USIC, University of Delhi) and Prof. P.K. Das, IISc Bangalore, India for SHG measurements. AS thanks DST-SERB for fellowship.

We are very grateful to Dr V.B. Nalbandyan for providing the samples for study of physical properties and for many useful critical comments on the manuscript. Support by Russian Foundation for Basic Research through grant 18-02-00326 is gratefully acknowledged. Specific heat measurements and theoretical simulations of the magnetic susceptibility were supported by the Russian Scientific Foundation via program RSF 17-12-01207 for EZ and SS. The work at KHU was supported by Basic Science Research Program through the National Research Foundation of Korea (NRF) funded by the Ministry of Education (NRF-2017R1D1A1B03029624). Mössbauer studies were supported by the Russian Science Foundation via program RSF 19-73-10034 for IG and AS. This work has been supported by the Ministry of Education and Science of the Russian Federation in the framework of Increase Competitiveness Program of NUST "MISiS" Grant No. K2-2017-084; by Act 211 of the Government of Russian Federation, Contracts No. 02.A03.21.0004 and No. 02.A03.21.0011.